\documentclass[draft]{agujournal2019}
\usepackage{url} %this package should fix any errors with URLs in refs.
\usepackage{lineno}
\usepackage[finalnew]{trackchanges}
\usepackage{soul}

\usepackage{color}
\usepackage{amsmath}
\usepackage{comment}
\usepackage{amssymb}

\draftfalse

\newcommand{\planss} {{Planetary Space Science }}  %%Planetary
\newcommand{\ssr}{   {Space Sci. Rev. }}

\newcommand{\prl}{{Phys. Rev. Lett.}}
\newcommand{\jgr}{   {J. Geophys. Res.}}
\newcommand{\grl}{   {Geophys. Res. Lett.}}
\newcommand{\apj}{   {Astrophys. J.}}

\graphicspath{{./}{figures/}}

\journalname{JGR: Space Physics}

\begin{document}

%% ---------------------------------------------------------------%%

\title{Nonresonant scattering of energetic electrons by electromagnetic ion cyclotron waves: spacecraft observations and theoretical framework}

\authors{Xin An\affil{1}, Anton Artemyev\affil{1}, Vassilis Angelopoulos\affil{1}, Xiao-Jia Zhang\affil{2}, Didier Mourenas\affil{3,4}, Jacob Bortnik\affil{5}, Xiaofei Shi\affil{1}} 
\affiliation{1}{Department of Earth, Planetary, and Space Sciences, University of California, Los Angeles, Los Angeles, CA, 90095, USA}
%\affiliation{2}{Space Research Institute of the Russian Academy of Sciences, Moscow, 117997, Russia}
\affiliation{2}{Department of Physics, University of Texas at Dallas, Richardson, TX, 75080, USA.}
\affiliation{3}{CEA, DAM, DIF, Arpajon, 91297, France}
\affiliation{4}{Laboratoire Mati\`ere en Conditions Extr\^emes, Paris-Saclay University, CEA, Bruy\`eres-le-Ch\^atel, 91190, France}
\affiliation{5}{Department of Atmospheric and Oceanic Sciences, University of California, Los Angeles, Los Angeles, CA, 90095, USA}

\correspondingauthor{Xin An}{phyax@ucla.edu}

\begin{keypoints}
\item {The theoretical model of nonresonant scattering is verified for \remove{realistic }wave packets derived from self-consistent simulations.}
\item {\remove{Realistic }Short EMIC wave packets extend the energies of efficient scattering well below the minimum resonance energy, consistent with the theory.}
%\item {EMIC wave power spectra are inferred from statistical ELFIN observations of relativistic electron precipitation, including nonresonant scattering.}
\item {EMIC wave power spectra are inferred from ELFIN observations of relativistic electron precipitation, including nonresonant scattering.}
\end{keypoints}

\begin{abstract}
Electromagnetic ion cyclotron (EMIC) waves lead to rapid scattering of relativistic electrons in Earth's radiation belts, due to their large amplitudes relative to other waves that interact with electrons of this energy range. A central feature of electron precipitation driven by EMIC waves is deeply elusive. That is, moderate precipitating fluxes at energies below the minimum resonance energy of EMIC waves occur concurrently with strong precipitating fluxes at resonance energies in low-altitude spacecraft observations. This paper expands on a previously reported solution to this problem: nonresonant scattering due to wave packets\remove{ of finite size}. The quasi-linear diffusion model is generalized to incorporate nonresonant scattering by a generic wave shape. The diffusion rate decays exponentially away from the resonance, where shorter packets lower decay rates and thus widen the energy range of significant scattering. Using realistic EMIC wave packets from $\delta f$ particle-in-cell simulations, test particle simulations are performed to demonstrate that intense, short packets extend the energy of significant scattering well below the minimum resonance energy, consistent with our theoretical prediction. Finally, \remove{we compare }the calculated precipitating-to-trapped flux ratio of relativistic electrons is compared to ELFIN observations, and the wave power spectra is inferred based on the measured flux ratio. We demonstrate that even with a narrow wave spectrum, short EMIC wave packets can provide moderately intense precipitating fluxes well below the minimum resonance energy.
\end{abstract}

\section*{Plain Language Summary}
Electromagnetic ion cyclotron (EMIC) waves are one of the most important plasma emissions in the near-Earth space. When electrons experience an approximately constant EMIC wave phase in gyration, they resonate with these waves and are scattered to precipitate to the Earth's upper atmosphere. Such cyclotron resonance between electrons and EMIC waves are typically above $1$\,MeV of electron energy. However, spacecraft at low Earth orbit often observe that electrons in the hundreds of keV range, which are not in resonance with EMIC waves, precipitate simultaneous with those $>1$\,MeV. Strongly modulated EMIC wave packets are promising in precipitating the sub-MeV electrons through nonresonant interactions. Here, the theoretical model of nonresonant scattering is verified for realistic EMIC wave packets from self-consistent computer simulations. EMIC wave power spectra are inferred from electron precipitation measurements by ELFIN. Short EMIC wave packets are shown to give a better agreement between the theoretical and observed precipitating-to-trapped flux ratios.

\section{Introduction}\label{sec:intro}

% \toAnton{i think i can write some extended intro starting from original quasi-linear papers \cite{Drummond&Pines62,Vedenov62} and to problem of nonresonant effect inclusion}
Relativistic electrons in Earth's radiation belts, at energies from hundreds of keV to several MeV, pose a significant threat to spacecraft and astronauts, by causing deep dielectric charging and high levels of radiation dose \cite{horne2021satellite,hands2018radiation}. Such electrons can also be scattered into the loss cone to penetrate down to the ionosphere and upper atmosphere, altering the ionospheric conductance \cite{robinson1987calculating,ridley2004ionospheric,khazanov2018impact} or driving loss in the ozone layer \cite{thorne1980importance,rozanov2012influence,seppala2015substorm}. The dynamic variation of the relativistic electron flux is controlled by a complex interplay between acceleration, transport, and loss processes \cite{Shprits08:JASTP_local,Li&Hudson19,Thorne10:GRL}. Despite other loss mechanisms \cite<e.g., magnetopause shadowing at the dayside, field line curvature scattering at the nightside; see>[]{Turner12,Sorathia18,Sergeev&Tsyganenko82,Artemyev13:angeo:scattering}, the focus of this paper is on the precipitation of relativistic electrons caused by electromagnetic ion cyclotron (EMIC) waves \cite{Thorne&Kennel71,Shoji&Omura12:precipitation,Blum15:precipitation, Usanova14, Shprits17, Kubota&Omura17, Grach&Demekhov20}. Because EMIC wave amplitudes, ranging between $0.1$--$10$\,nT \cite{Min12,Zhang16:grl}, are considerably larger than other waves (e.g., whistler-mode waves) that can interact with relativistic electrons, scattering can be very rapid, especially during the main phase of a geomagnetic storm, when the waves are intensified. The large-amplitude ($>1$\,nT) EMIC waves are typically observed during geomagnetic active times in the dayside outer magnetosphere and dusk-to-noon inner magnetosphere \cite{Keika13,Usanova12,Zhang16:grl,ross2021variability}.

\begin{figure}
    \centering
    \includegraphics[width=0.6\textwidth]{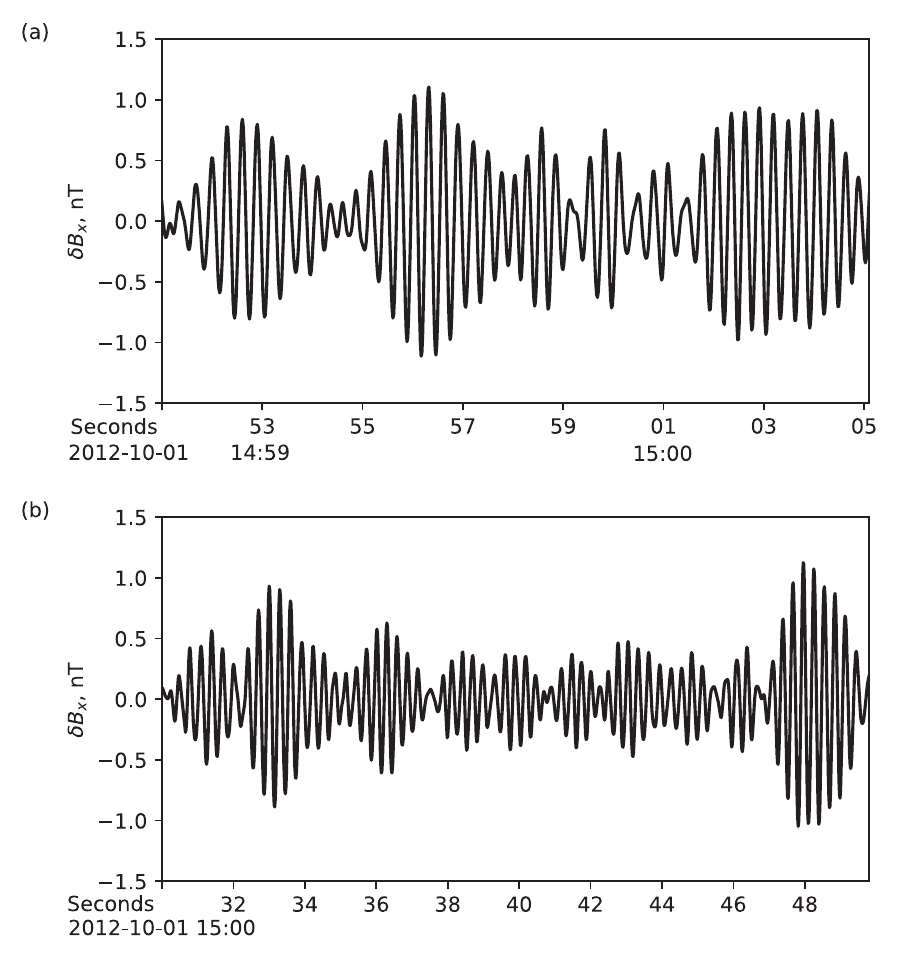}
    \caption{\add{Examples of EMIC wave packets observed by Van Allen Probe A. Panels (a) and (b) show relatively long and short wave packets, respectively. The field-aligned coordinate system is used. The $z$ direction is defined along the background magnetic field. The $y$ direction is defined along the cross product of $z$ and Geocentric Solar Ecliptic (GSE) $x$. The $x$ direction completes the Cartesian coordinate system. The shown magnetic field component is in the $x$ direction of the field-aligned coordinate.}}
    \label{fig:emic-van-allen-probes}
\end{figure}

EMIC waves are left-hand polarized plasma emissions below the proton gyrofrequency in the Earth's magnetosphere. They are excited by the pitch-angle anisotropy of ring current ions ($T_\perp > T_\parallel$; $T_\perp$ and $T_\parallel$ being perpendicular and parallel temperatures, respectively) through the cyclotron resonant instability \cite<e.g.,>[]{Kennel&Petschek66,Shoji&Omura13:generation,Min15}. This enhanced anistropy can be provided by injections from the plasma sheet \cite{Chen10:emic,remya2020association,jun2021characteristics,yahnin2021evening}, by solar wind dynamic pressure increases \cite{chen2020statistical,xue2021prompt,Jun19:emic,yahnin2019simultaneous}, and by ultra-low-frequency wave modulation \cite{rasinkangas1998modulation,loto2009modulation}. EMIC waves are often observed to be amplitude modulated and appear as strong, short packets \cite<having a peak amplitude comparable to or larger than $1$\,nT and a few wave periods in each packet; e.g.,>[]{Jacobs64,obayashi1965hydromagnetic,perraut1984ion,fraser1985observations,fraser2006electromagnetic,Usanova10,Shoji&Omura18:emic_packets,An22:prl,Grach21:emic}.\add{ Two examples of such EMIC wave packets observed the fluxgate magnetometer }\cite{Kletzing13}\add{ on Van Allen Probe A are shown in Figure }\ref{fig:emic-van-allen-probes}\add{.} The modulation of EMIC waves may be caused by ion cyclotron trapping \cite<see>[]{Shoji&Omura17:emic_trapping} during the excitation of EMIC waves \cite<see discussions in>[for analogous modulations of whistler and Langmuir waves]{Tao17:generation,Trakhtengerts04,ONeil65}, by multi-frequency interference, or even by ultra-low frequency waves \cite<e.g.,>[]{liu2019magnetospheric}.

To resonate with the left-hand polarized EMIC waves, electrons must overtake the wave so that the wave polarization is reversed in the electron frame and the Doppler-shifted wave frequency matches the electron gyrofrequency. This is only possible above a certain minimum resonance energy (usually $\geq 1$\,MeV) \cite{Summers&Thorne03,Ni15}. However, detailed comparisons between low-altitude spacecraft observations and theoretical predictions of precipitating electrons reveal a significant discrepancy. That is, electrons in the hundreds of keV energy range, which are well below the minimum resonance energy of EMIC waves, are often observed to precipitate simultaneously with those at $\geq 1$\,MeV \cite{Yahnin16,Hendry17,Capannolo19:microburst}. This discrepancy cannot be resolved by the inclusion of hot plasma dispersion relation of EMIC waves \cite{cao2017scattering,Chen19}. Nonresonant scattering by wave packets of finite size, on the other hand, is a promising mechanism to extend the energy range of efficient scattering well below the minimum resonance energy \cite{Chen16:nonresonant,An22:prl}. The main idea is that electrons experience a net change in magnetic moment (i.e., the first adiabatic invariant) when the variation of wave amplitude is significant in one wave length, even if electrons are away from the resonance. Such change in the electron magnetic moment is accumulated over a certain number of wave lengths, and thus nonresonant electrons can be scattered. Equivalently, fast variations of wave amplitude introduce a spread of power in wavenumber space, extending the ``effective'' resonance energy.

Nonresonant interactions have applications in a wide range of contexts. For example, electron interactions with intense, localized Langmuir wave packets involve transit-time (nonresonant) scattering \cite{goldman1984strong,robinson1997nonlinear}, which occur in the auroral ionosphere \cite{muschietti1994interaction}, solar wind \cite{Krafft13,rowland1977simulations,gurnett1976electron}, planetary bow shocks \cite{kellogg2003langmuir,anderson1981plasma,gurnett1981parametric}, ionospheric-modification experiments \cite{dubois1993excitation}, electron-beam experiments \cite{leung1982plasma,sun2022electron}, and laser-plasma experiments \cite{rubenchik1991strong}. In addition, electron interactions with time domain structures (i.e., electric field spikes) are a case of nonresonant scattering, occurring in planetary magnetospheres and auroral ionosphere \cite{Mozer15,Vasko17:diffusion}. Such nonresonant effects naturally complement the classical quasi-linear theory of resonant diffusive scattering \cite{Vedenov62,Drummond&Pines62,Andronov&Trakhtengerts64,Kennel&Engelmann66}, and provide an effective extension of scattering rates below resonance energies \cite{An22:prl}. The aim of this study is to provide the full theoretical basis and tool for accounting for nonresonant electron scattering by EMIC waves in radiation belt models. % Moreover, driving plasma currents using radio-frequency waves of finite packet sizes can be achieved via nonresonant so-called ``ponderomotive'' forces, relevant to fusion devices \cite{lamb1984behavior,fisch2003current,dodin2005nonlinear}.

This paper is organized as follows. In Section \ref{sec:theory}, a detailed description of how to incorporate nonresonant interactions into the quasi-linear diffusion framework is provided. In Section \ref{sec:modulation}, the theoretical predictions are verified against electron scattering by realistic EMIC wave packets from $\delta f$-PIC simulations. In Section \ref{sec:obserations}, magnetic power spectra are inferred from statistical ELFIN observations of relativistic electron precipitation. This study is summarized and concluded in Section \ref{sec:conclusions}.

\section{Theoretical model of nonresonant wave-particle interactions}\label{sec:theory}
Most of the materials presented in this section can be found in two publications by \citeA{An22:prl} and \citeA{grach2023interaction}. However, we prefer to rederive the main equations and supplement them with additional explanations to provide self-contained derivations and model verification.

%\toXin{i think we can provide quite strict theoretical description here... may be with some schematic figure. This is better to be done for gaussina+sin wave}
\subsection{Equation of motion}
The equation of motion for an electron moving through an EMIC wave packet propagating along the geomagnetic field line is
\begin{linenomath*}
\begin{equation}\label{eq:eq-of-motion-vec}
	\frac{\mathrm{d} \mathbf{u}}{\mathrm{d} t} = -\frac{e}{m} \left[\delta \mathbf{E} + \frac{\mathbf{u} \times \left(\mathbf{B}_0 + \delta \mathbf{B}\right)}{\gamma c}\right] ,
\end{equation}
\end{linenomath*}
where $t$ is time, $\mathbf{u}$ is the relativistic velocity, $-e$ and $m$ are the charge and mass of the electron, $c$ is the speed of light, $\gamma$ is the Lorentz factor, $\delta \mathbf{E}$ and $\delta \mathbf{B}$ are the electric and magnetic fields of the wave packet, and $\mathbf{B}_0$ is the Earth's dipole magnetic field. The EMIC wave fields of frequency $\omega$, wavenumber $k$ and magnetic amplitude $B_w$ are
\begin{linenomath*}
\begin{equation}\label{eq:wave-fld}
	\begin{split}
		&\delta B_x = B_w g(z) \cos(\phi) , \hspace{15pt}
		\delta B_y = B_w g(z) \sin(\phi) , \hspace{15pt}
		\delta B_z = 0, \\
		&\delta E_x = \frac{\omega}{k c} \delta B_y , \hspace{15pt}
		\delta E_y = -\frac{\omega}{k c} \delta B_x , \hspace{15pt}
		\delta E_z = 0,
	\end{split}
\end{equation}
\end{linenomath*}
where $z$ is the field-aligned coordinate, $x$ and $y$ are the two perpendicular coordinates, $\phi = \int \left(k \mathrm{d}z - \omega \mathrm{d}t\right)$ is the wave phase, and $g(z)$ is the spatial wave shape function. Note that $k$ is the main wavenumber of the EMIC wave packet. \add{The relationships between $\mathbf{E}$ and $\mathbf{B}$ in Equation }\eqref{eq:wave-fld}\add{ are approximate because of the plane wave assumption.} A reduced dipole field model \cite{Bell84} is adopted to describe the geomagnetic field $\mathbf{B}_0$:
\begin{linenomath*}
\begin{equation}\label{eq:background-B-fld}
    B_{0z} = B_{\mathrm{eq}} (1 + \xi z^2), \hspace{15pt} B_{0x} = -\frac{x}{2} \frac{\partial B_{0z}}{\partial z}, \hspace{15pt} B_{0y} = -\frac{y}{2} \frac{\partial B_{0z}}{\partial z},
\end{equation}
\end{linenomath*}
where $B_{\mathrm{eq}}$ is the magnetic field at the equator, $\xi = 9 / (2 L^2 R_E^2)$, $L$ is the $\mathrm{L}$-shell number, and $R_E$ is the Earth's radius. Along electron gyro-orbits, the two perpendicular coordinates in Equation \eqref{eq:background-B-fld} are given by $x = u_y / \omega_{ce}$ and $y = -u_x / \omega_{ce}$, where $\omega_{ce}(z) = e B_{0z}(z) / mc$ is the electron gyrofrequency. It is assumed that the particle gyroradius is much less than $L R_E$ in this reduced dipole field model. Plugging the wave fields [Equation \eqref{eq:wave-fld}] and the background magnetic field [Equation \eqref{eq:background-B-fld}] into Equation \eqref{eq:eq-of-motion-vec}, we obtain
\begin{linenomath*}
\begin{equation}\label{eq:motion-cartesian}
\begin{split}
    \frac{\mathrm{d} u_x}{\mathrm{d} t} &= -\frac{e}{m} \left[-\left(\frac{u_z}{\gamma c} - \frac{\omega}{kc}\right) \delta B_y + \frac{u_y}{\gamma c} B_{0z} - \frac{u_x u_z}{2 \gamma c \omega_{ce}} \frac{\partial B_{0z}}{\partial z}\right] , \\
	\frac{\mathrm{d} u_y}{\mathrm{d} t} &= -\frac{e}{m} \left[\left(\frac{u_z}{\gamma c}  - \frac{\omega}{kc}\right) \delta B_x - \frac{u_x}{\gamma c} B_{0z} - \frac{u_y u_z}{2 \gamma c \omega_{ce}} \frac{\partial B_{0z}}{\partial z}\right] , \\
	\frac{\mathrm{d} u_z}{\mathrm{d} t} &= -\frac{e}{m} \left[\frac{u_x \delta B_y - u_y \delta B_x}{\gamma c} + \frac{u_x^2 + u_y^2}{2 \gamma c \omega_{ce}} \frac{\partial B_{0z}}{\partial z}\right] .
\end{split}
\end{equation}
\end{linenomath*}
Velocities $(u_x, u_y, u_z)$ are transformed to new variables $(I, \varphi, \gamma)$, where $I = m(u_x^2 + u_y^2) / (2 \omega_{ce})$ is the electron magnetic moment, $\varphi = \phi - \arctan(u_y / u_x)$ is the phase angle between the particle perpendicular velocity and the wave magnetic field, and the Lorentz factor is $\gamma = \sqrt{1 + (u_z/c)^2 + 2 \omega_{ce} I / (m c^2)}$. The equation of motion can rewritten as
\begin{linenomath*}
\begin{eqnarray}
	\frac{\mathrm{d} I}{\mathrm{d} t} &=& \left(\frac{u_z}{\gamma} - \frac{\omega}{k}\right) \sqrt{\frac{2 I}{mc^2 \omega_{ce}}} e B_w g(z) \sin\varphi ,\\
	%\frac{\mathrm{d} u_z}{\mathrm{d} t} = -\frac{\omega_{ce,\mathrm{eq}} c}{\gamma} \sqrt{\frac{2 \omega_{ce} I}{m c^2}} \frac{B_w}{B_{\mathrm{eq}}} g(z) \sin\varphi - \frac{I}{\gamma m} \frac{\partial \omega_{ce}}{\partial z} ,\\
	\frac{\mathrm{d} \varphi}{\mathrm{d} t} &=& \frac{k u_z}{\gamma} - \omega - \frac{\omega_{ce}}{\gamma} +\omega_{ce,\mathrm{eq}} \sqrt{\frac{m c^2}{2 \omega_{ce} I}} \left(\frac{u_z}{\gamma c} - \frac{\omega}{k c}\right) \frac{B_w}{B_{\mathrm{eq}}} g(z) \cos\varphi ,\label{eq:dvarphidt}\\
	\frac{\mathrm{d} \gamma}{\mathrm{d} t} &=& -\omega_{ce,\mathrm{eq}} \frac{\omega}{\gamma k c} \sqrt{\frac{2 \omega_{ce} I}{m c^2}} \frac{B_w}{B_{\mathrm{eq}}} g(z) \sin\varphi ,
\end{eqnarray}
\end{linenomath*}
where $u_z = \gamma \mathrm{d} z / \mathrm{d} t = c \sqrt{\gamma^2 - 1 - 2 \omega_{ce} I / (m c^2)}$. Because the wave phase velocity is much smaller than the electron velocity near gyroresonance ($\omega / k \ll u_z / \gamma$), electrons are dominantly scattered in pitch angle while their energies are approximately constant ($\gamma = \mathrm{constant}$). This is equivalent to neglecting electric field. Furthermore, the last term in Equation \eqref{eq:dvarphidt} is only important for very small pitch angles, which can be neglected for our application. It will be convenient to use $z$ as the independent variable. Thus the equation of motion is simplified as
\begin{linenomath*}
\begin{eqnarray}
    \frac{\mathrm{d} I}{\mathrm{d} z} &=& \sqrt{\frac{2 I}{mc^2 \omega_{ce}(z)}} e B_w g(z) \sin\varphi ,\\
    \frac{\mathrm{d} \varphi}{\mathrm{d} z} &=& k - \frac{\omega_{ce}(z) / c}{\sqrt{\gamma^2 - 1 - 2 \omega_{ce}(z) I / (m c^2)}} .
\end{eqnarray}
\end{linenomath*}

\subsection{Perturbation analysis}
For resonant interactions with large-amplitude EMIC waves, the perturbation analysis with the assumption of small-amplitude waves is invalid. However, for nonresonant interactions, the particle orbits do not deviate much from the zeroth-order gyro-orbits even with large-amplitude waves, making the perturbation analysis appropriate. In addition, short EMIC wave packets can disrupt nonlinear resonance effects, causing the the wave-particle resonant interaction to revert to a classical, diffusive scattering regime. For interactions with the modulated wave field, using the perturbation method, the first-order change of the electron magnetic moment is obtained by integrating $\mathrm{d} I / \mathrm{d} z$ along the zeroth-order orbit
\begin{linenomath*}
\begin{equation}
    \Delta I = \int_{z_l}^{z_u}\mathrm{d}z' \sqrt{\frac{2 I_0}{mc^2 \omega_{ce}(z')}} e B_w g(z') \sin\left(\varphi(z')\right) ,
\end{equation}
\end{linenomath*}
%and the second-order change of the electron magnetic moment by integrating $\mathrm{d} I / \mathrm{d} z$ along the first-order orbit
%\begin{linenomath}
%\begin{equation}
%    \Delta I^{(2)} = \int_{z_l}^{z_u}\mathrm{d}z' \sqrt{\frac{2 \left(I^{(0)} + \Delta I^{(1)}\right)}{mc^2 \omega_{ce}(z')}} e B_w g(z') \sin\left(\varphi^{(0)}(z') + \Delta\varphi^{(1)}(z')\right),
%\end{equation}
%\end{linenomath}
where the phase angle is
\begin{linenomath*}
\begin{equation}
    \varphi(z) = \varphi_0 + \int_{z_l}^{z}\mathrm{d}z' \left(k - \frac{\omega_{ce}(z') / c}{\sqrt{\gamma^2 - 1 - 2 \omega_{ce}(z') I_0 / (m c^2)}}\right) = \varphi_0 + \varphi_R(z) ,
\end{equation}
\end{linenomath*}
$z_l$ and $z_u$ are the lower and upper boundaries of the wave packet, respectively, $I_0$ is the zeroth-order electron magnetic moment, and $\varphi_R(z)$ denotes the integral.

The variance of electron magnetic moment is
\begin{linenomath*}
\begin{equation}\label{eq:ensemble-avg-dI2}
\begin{split}
    \left\langle (\Delta I)^2 \right\rangle &= \frac{2 I_0 e^2 B_w^2}{m c^2} \int_{z_l}^{z_u}\mathrm{d}z' \int_{z_l}^{z_u}\mathrm{d}z'' \frac{g(z') g(z'')}{\sqrt{\omega_{ce}(z') \omega_{ce}(z'')}} \left\langle \sin\varphi(z') \sin\varphi(z'') \right\rangle \\
    &= \frac{I_0 e^2 B_w^2}{m c^2} \int_{z_l}^{z_u}\mathrm{d}z' \int_{z_l}^{z_u}\mathrm{d}z'' \frac{g(z') g(z'')}{\sqrt{\omega_{ce}(z') \omega_{ce}(z'')}} \cos\left[\varphi_R(z') - \varphi_R(z'')\right] \\
    &= \frac{I_0 e^2 B_w^2}{m c^2 \omega_{ce}(z_c)} \left\vert \int_{z_l}^{z_u} \mathrm{d} z' g(z') e^{i \varphi_R(z')} \right\vert^2 ,
\end{split}
\end{equation}
\end{linenomath*}
where $\langle \cdot \rangle$ represents the ensemble average over $\varphi_0$, and $z_c$ is the center of the wave packet. Because $\omega_{ce}$ varies with $z$ on the scale $\xi^{-\frac{1}{2}}$ ($\sim L R_E$), which is large compared to the size of those wave packets, $\omega_{ce} (z')$ is approximated as $\omega_{ce} (z_c)$ and is taken it out of the integral.

The scattering factor is defined as \cite{An22:prl,grach2023interaction}
\begin{linenomath*}
\begin{equation}\label{eq:G-exact}
    G = \left\vert \int_{z_l}^{z_u} \mathrm{d} z' g(z') e^{i \varphi_R(z')} \right\vert^2 
\end{equation}
\end{linenomath*}
that controls the electron scattering by a wave packet. This quantity might be nonzero under two scenarios. The first scenario occurs when the shape function $g(z)$ changes rapidly (i.e., peak-to-peak wave amplitude changes significantly in one cycle) so that the electron magnetic moment has a net change in one wave period \cite{Chen16:nonresonant,An22:prl,grach2023interaction}. Such changes in the electron magnetic moment accumulate over a certain interaction length, leading to nonresonant scattering. Some examples include electron scattering by equatorially confined magnetosonic waves \cite{Bortnik&Thorne10,Bortnik15} and by time domain structures around injection fronts \cite{Vasko17:diffusion}. The second scenario occurs when the phase angle $\varphi_R$ stays almost stationary near the resonance $\mathrm{d}\varphi_R / \mathrm{d}z = 0$, leading to resonant scattering. The electrons of interest in this study are those that do not have sufficiently high energies to resonate with EMIC waves, but may still be scattered if they are close to the resonance, causing both effects described above to have a contribution to the net scattering.

\subsection{Resonance condition}
The wave shape function is Fourier analyzed in wavenumber space, $g(z) = \int_{-\infty}^{\infty} \mathrm{d}\kappa\,\, \hat{g}(\kappa) e^{i \kappa z}$, where $\kappa$ is the wavenumber. The scattering factor is thus rewritten as
\begin{linenomath*}
\begin{equation}
    G = \left\vert \int_{-\infty}^{\infty}\mathrm{d}\kappa\,\, \hat{g}(\kappa) \int_{\Psi_l}^{\Psi_u} \frac{\mathrm{d}\Psi}{\dot{\Psi}(z)} e^{i \Psi} \right\vert^2 ,
\end{equation}
\end{linenomath*}
where $\Psi(z) = \kappa z + \varphi_R (z)$, $\Psi_l = \Psi(z_l)$, and $\Psi_u = \Psi(z_u)$. The phase angle $\Psi$ varies rapidly except in the vicinity of the singularity $\dot{\Psi}(z) = \mathrm{d} \Psi / \mathrm{d} z = 0$, i.e., the resonance condition, where the phase integral is nonzero. The resonance location $z_0$ is determined by
\begin{linenomath*}
\begin{equation}\label{eq:resonance-condition}
    \dot{\Psi}\big\vert_{z = z_0} = \kappa + k - \frac{\omega_{ce}(z_0)}{c \sqrt{\gamma^2 - 1 - 2 \omega_{ce}(z_0) I_0 / (m c^2)}} = 0 .
\end{equation}
\end{linenomath*}
By solving this resonance condition, we obtain
\begin{linenomath*}
\begin{equation}
	z_0 = 
	\begin{cases}
		\pm \sqrt{\left(\frac{\omega_{ce, R}}{\omega_{ce, \mathrm{eq}}} - 1\right) / \xi} & \text{for } \omega_{ce,R} \geqslant \omega_{ce, \mathrm{eq}} \text{ (resonant)}  , \\
		\pm i \sqrt{\left(1 - \frac{\omega_{ce, R}}{\omega_{ce, \mathrm{eq}}}\right) / \xi} & \text{for } \omega_{ce,R} < \omega_{ce, \mathrm{eq}} \text{ (nonresonant)} ,
	\end{cases}
\end{equation}
\end{linenomath*}
where
\begin{linenomath*}
\begin{equation}
    \omega_{ce, R} = (\kappa + k)^2 c^2 \left[-(I_0 / m c^2) + \sqrt{(I_0 / m c^2)^2 + (\gamma^2 - 1) (\kappa + k)^{-2} c^{-2}}\right] .
\end{equation}
\end{linenomath*}
The resonance location is on the real $z$-axis for resonant electrons, whereas it moves to the complex $z$ plane for nonresonant electrons.

It is useful to map the resonance location from $z_0$ to the phase angle $\Psi_0$:
\begin{linenomath*}
\begin{equation}
    \begin{split}
        \Psi_0 &= \int_{z_l}^{z_0} \mathrm{d}z \left(\kappa + k - \frac{\omega_{ce}(z)}{c \sqrt{\gamma^2 - 1 - 2\omega_{ce}(z) I_0 / (m c^2)}}\right) \\
        &= \left(\int_{z_l}^{\operatorname{Re}(z_0)} \mathrm{d}z + \int_{\operatorname{Re}(z_0)}^{z_0} \mathrm{d}z\right) \left(\kappa + k - \frac{\omega_{ce}(z)}{c \sqrt{\gamma^2 - 1 - 2\omega_{ce}(z) I_0 / (m c^2)}}\right) \\
        &= \operatorname{Re}(\Psi_0) + i \operatorname{Im}(\Psi_0) .
    \end{split}
\end{equation}
\end{linenomath*}
For resonant interactions, it suffices to know that $\Psi_0$ is a real number. For nonreosnant interactions, the imaginary part of phase angle can be calculated explicitly:
\begin{linenomath*}
\begin{equation}
\begin{split}
    \operatorname{Im}(\Psi_0) &= \frac{1}{2 \sqrt{\xi}} \left\{ -\left(1 - \frac{\omega_{ce, R}}{\omega_{ce, \mathrm{eq}}}\right)^{1/2} \left[\frac{3 (\kappa + k)}{2} + \frac{(\kappa + k)}{2}\left(1 + \frac{\gamma^2 - 1}{(\kappa + k)^2 (I_0/mc)^2}\right)^{1/2}\right] \right. \\
		&\left. + \frac{(\gamma^2 -1) m c^2 + 2 I_0 \omega_{ce, \mathrm{eq}}}{(2 I_0)^{3/2} (\omega_{ce, \mathrm{eq}} / m)^{1/2}} \ln\left[\frac{\left(\frac{(\gamma^2 - 1) m c^2}{2 I_0 \omega_{ce, \mathrm{eq}}} - 1\right)^{1/2}}{\left(\frac{(\gamma^2 - 1) m c^2}{2 I_0 \omega_{ce, \mathrm{eq}}} - \frac{\omega_{ce, R}}{\omega_{ce, \mathrm{eq}}}\right)^{1/2} - \left(1 - \frac{\omega_{ce, R}}{\omega_{ce, \mathrm{eq}}}\right)^{1/2}}\right] \right\} ,
\end{split}
\end{equation}
\end{linenomath*}
which will be useful later in the evaluation of scattering rates.

\subsection{Resonant and nonresonant regimes}
\begin{figure}[tphb]
    \centering
    \includegraphics[width=\textwidth]{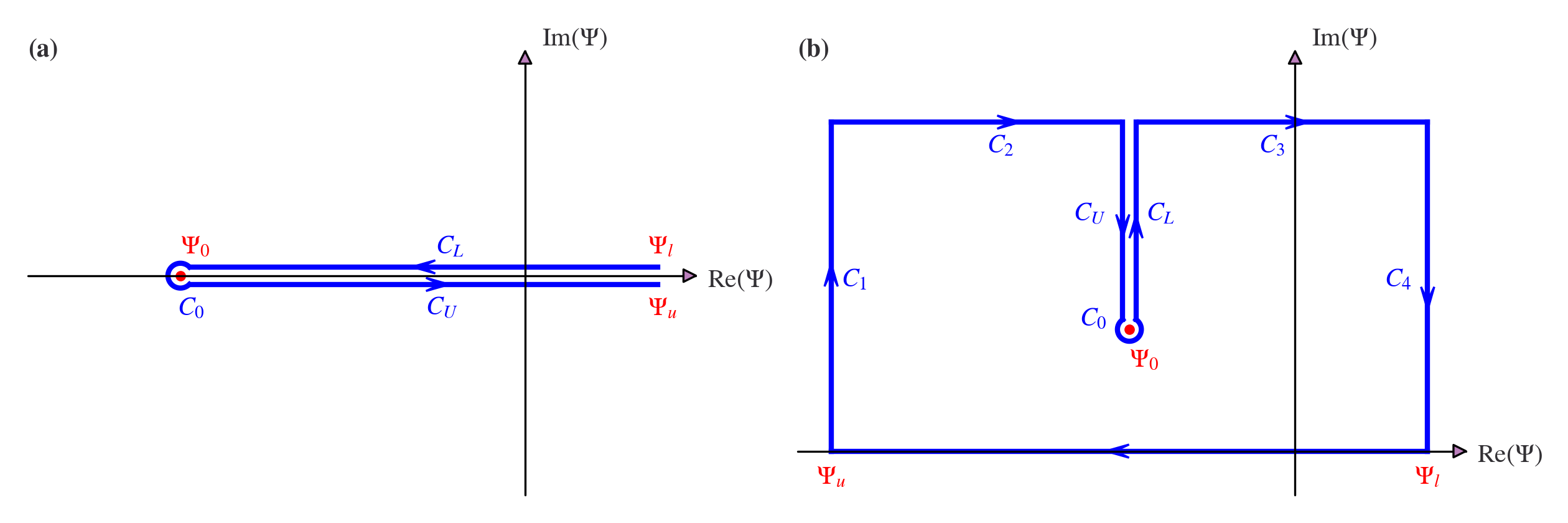}
    \caption{Two regimes of resonance locations in $(\operatorname{Re}(\Psi), \operatorname{Im}(\Psi))$ and the associated contours of phase integral in Equation \eqref{eq:phase-integral}. (a) Resonant regime. The resonance location $\Psi_0$ is on the real axis. The phase integral in Equation \eqref{eq:phase-integral} extends from $(\Psi_l, \Psi_0 + \delta)$, around $\Psi_0$ counter-clockwise and back to $(\Psi_0 + \delta, \Psi_u)$, where $\delta$ is a small positive number. Such integration path is called the Hankel contour \cite{krantz1999handbook}. (b) Nonresonant regime. $\Psi_0$ is in the upper half plane. The phase integral in Equation \eqref{eq:phase-integral} extending along the real-$\Psi$ axis from $\Psi_l$ to $\Psi_u$ is deformed into the upper half of the plane.}
    \label{fig:integral-contours}
\end{figure}

Most of the contribution to the scattering factor $G$ comes from the vicinity of the resonance location $z_0$. $\Psi(z)$ and $\dot{\Psi}(z)$ are expanded as Taylor series about $z = z_0$:
\begin{linenomath*}
\begin{eqnarray}
    \Psi(z) &=& \Psi_0 + \frac{1}{2} \Ddot{\Psi}_0 (z - z_0)^2 , \\
    \dot{\Psi}(z) &=& \Ddot{\Psi}_0 (z - z_0) ,
\end{eqnarray}
\end{linenomath*}
where
\begin{linenomath*}
\begin{equation}\label{eq:ddot-psi0}
    \Ddot{\Psi}_0 = \frac{\mathrm{d}^2 \Psi}{\mathrm{d}z^2} (z_0) = -\frac{1}{c} \frac{\mathrm{d} \omega_{ce} (z_0)}{\mathrm{d} z} \left[\gamma^2 - 1 - \frac{\omega_{ce}(z_0) I_0}{m c^2}\right] \left[\gamma^2 - 1 - \frac{2 \omega_{ce}(z_0) I_0}{m c^2}\right]^{-\frac{3}{2}} .
\end{equation}
\end{linenomath*}
Using these Taylor expansions, $\dot{\Psi}(z)$ may be expressed in terms of $\Psi$ as
\begin{linenomath*}
\begin{equation}
    \dot{\Psi}(z) = \left[2 \ddot{\Psi}_0 \left(\Psi - \Psi_0\right)\right]^{\frac{1}{2}} .
\end{equation}
\end{linenomath*}
The phase integral inside the scattering factor $G$ is thus written as
\begin{linenomath*}
\begin{equation}\label{eq:phase-integral}
    \int_{\Psi_l}^{\Psi_u} \frac{\mathrm{d}\Psi}{\dot{\Psi}(z)} e^{i \Psi} = \frac{1}{(2 \ddot{\Psi}_0)^{\frac{1}{2}}} e^{i \Psi_0} 	\int_{\Psi_l}^{\Psi_u} \frac{\mathrm{d}\Psi}{(\Psi - \Psi_0)^{\frac{1}{2}}} e^{i (\Psi - \Psi_0)} .
\end{equation}
\end{linenomath*}
Because the fractional power of $\Psi - \Psi_0$ occurs in the denominator, $\Psi_0$ is a branch point such that the contour of this integral should go through an arbitrarily small circuit around $\Psi_0$. Such contours are shown in Figure \ref{fig:integral-contours} for resonant and nonresonant interactions. The phase integral only survives on the branch cuts labeled $C_U$ and $C_L$ for either resonant or nonresonant regimes. In the latter regime, the phase integral can be evaluated as
\begin{linenomath*}
\begin{equation}\label{eq:phase_integral}
    \int_{\Psi_l}^{\Psi_u} \frac{\mathrm{d}\Psi}{\dot{\Psi}(z)} e^{i \Psi} = \left(\frac{2 \pi}{\ddot{\Psi}_0}\right)^{\frac{1}{2}} e^{i (\Psi_0 + \frac{\pi}{4})} ,
\end{equation}
\end{linenomath*}
which has the same form in the resonant case except for a trivial phase factor $-1$. The detailed calculation of the phase integral is given in \ref{append-calc-phase-integral}. It is noted that Equation \eqref{eq:phase-integral} has a singularity for the resonant interaction at the equator because of $\ddot{\Psi}_0(z_0 = 0) = 0$. The treatment of this interaction can be found in \citeA{grach2023interaction}.

Thus, for both resonant and nonresonant interactions, the scattering factor can be unified in one formula:
\begin{linenomath*}
\begin{equation}\label{eq:scattering-factor-approx}
    G = 2 \pi \left\vert \int_{-\infty}^{\infty}\mathrm{d}\kappa\,\, \hat{g}(\kappa) \frac{e^{i \Psi_0}}{\left(\ddot{\Psi}_0\right)^{\frac{1}{2}}} \right\vert^2 .
\end{equation}
\end{linenomath*}
In the limit of infinitely long wave packet $g(z) = 1$ associated with $\hat{g}(\kappa) = \delta(\kappa)$, we obtain
\begin{linenomath*}
\begin{equation}\label{eq:G-long-limit}
    G = 2 \pi \frac{e^{- 2 \operatorname{Im}(\Psi_0)}}{\vert \ddot{\Psi}_0 \vert} .
\end{equation}
\end{linenomath*}
For resonant interactions, we have $\operatorname{Im}(\Psi_0) = 0$ and the exponential factor $e^{- 2 \operatorname{Im}(\Psi_0)} = 1$. The scattering rate exponentially decays away from the resonance, and the decay rate is controlled by the imaginary part of the resonance location in the complex $\Psi$ plane. The denominator $\lvert \ddot{\Psi}_0 \rvert \propto \lvert \mathrm{d} \omega_{ce} (z_0) / \mathrm{d} z \rvert$ [see Equation \eqref{eq:ddot-psi0}] recovers the dependence of scattering rate on the inhomogeneity of background magnetic field in the narrowband limit \cite{Albert10}.

\subsection{Bounce-averaged pitch angle diffusion rate}
The equatorial pitch angle is related to the magnetic moment through
\begin{linenomath*}
\begin{equation}
	\sin^2 \alpha_{\mathrm{eq}} = \frac{2 \omega_{ce, \mathrm{eq}} I}{(\gamma^2 - 1) m c^2} . \label{eq:alpha-eq-I}
\end{equation}
\end{linenomath*}
The variance of the change of the equatorial pitch angle is given by
\begin{linenomath*}
\begin{equation}\label{eq:da2-dI2}
	\left\langle \left(\Delta \alpha_{\mathrm{eq}}\right)^2 \right\rangle = \frac{\omega_{ce,\mathrm{eq}} / (mc^2)}{2 I_0 \left[\gamma^2 - 1 - 2 \omega_{ce,\mathrm{eq}} I_0 / (mc^2)\right]} \left\langle \left(\Delta I\right)^2 \right\rangle .
\end{equation}    
\end{linenomath*}
Statistically, we let $N$ wave packets fill a field line over the time scale of one bounce period $\tau_b$. Each packet has an amplitude $\delta B_i$ and a scattering factor $G_i$ ($i = 1, 2, \cdots, N$). Based on Equations \eqref{eq:ensemble-avg-dI2} and \eqref{eq:da2-dI2}, the bounce-averaged pitch angle diffusion rate is
\begin{linenomath*}
\begin{equation}\label{eq:Daa-theory}
    D_{\alpha\alpha} = \frac{\langle(\Delta \alpha_{\mathrm{eq}})^2\rangle}{2 \tau_b} = \frac{e^2 \omega_{ce, \mathrm{eq}}}{4 (\gamma^2 - 1) m^2 c^4 \cos^2{\alpha_{\mathrm{eq}}} \omega_{ce}(z_c) \tau_b} \sum_{i = 1}^{N} \delta B_i^2 G_i .
\end{equation}
\end{linenomath*}

%\section{Test particle verification}\label{sec:testparticle}

%\toXin{i would suggest to provide here some verification for different wave parameters, but still for gaussina+sin wave}

\section{Effects of wave modulation on energetic electron scattering}\label{sec:modulation}

% \toXin{not sure about title of this section, but i would suggest to show several examples of EMIC wave packet observations (Xiajia?) + say that we would like to be sure we deal with real signal and thus show a couple examples of wave packets from pic. I this we can try to use the spectrum of such wave packets as an input parameter to exponential factor calculation (instead of gaussian+sin) and thus check that test particle and theory gives similar decay of scattering rates away from the resonance in realistic wave packets... alternatively, we may try to estimate the multiplication factor that would make theory for gaussian+sin looking similar to test particle simulations for realistic wave packets}

Depending on the wave shape function $g(z)$ of each wave packet, the energy range of efficient electron scattering can be greatly extended for short wave packets [see Equation \eqref{eq:resonance-condition} and discussions in, e.g., \citeA{An22:prl}]. Thus, it is important to use realistic wave packets to scatter electrons \cite{Grach21:emic}, in terms of both the wave shape and number of wave periods in each packet. To this end, PIC simulations are performed to generate EMIC wavepackets and then evaluate electron scattering rates by these EMIC waves using test particle simulations.

\subsection{Computational setup}
% basic setup of PIC simulations
The OSIRIS PIC framework \cite{fonseca2002osiris,fonseca2013exploiting} is extended here to simulate the excitation of EMIC waves in a simplified dipole magnetic field. The simulations have one dimension ($z$) in configuration space and three dimensions ($v_x$, $v_y$, $v_z$) in velocity space.\add{ Because the most efficient nonresonant scattering occurs close to the equator, where EMIC waves propagate parallel to the background magnetic field, the restriction of one-dimensional spatial domain does not affect the main results.} The computational domain spans $-163.84 \leq z / d_i \leq 163.84$ with a cell length $0.16\, d_i$, where $d_i = c / \omega_{pi}$ is the ion inertial length, and $\omega_{pi}$ is the reference ion plasma frequency. The background magnetic field is given by Equation \eqref{eq:background-B-fld} with $\xi = 7.05 \times 10^{-5}\, d_i^{-2}$, roughly corresponding to the magnetospheric location $L$-shell $7.5$ and the plasma density $3\, \mathrm{cm}^{-3}$. These magnetospheric conditions also \change{gives}{give} $\omega_{ce} / \omega_{pe} = 0.128$, where $\omega_{ce}$ is the electron gyrofrequency and $\omega_{pe}$ is the plasma frequency. Because the ion-to-electron mass ratio in the simulations is $100$, the normalized equatorial magnetic field is $\omega_{ci, \mathrm{eq}} / \omega_{pi} = 0.0128$. The \change{electromagnetic}{electric and magnetic} fields are advanced by integrating \change{Maxwell's Equations}{the Ampere's and Faraday's Laws, respectively,} using the leapfrog scheme. The particle position and velocity are updated using $\mathrm{d}z / \mathrm{d}t = v_z = u_z / \gamma$ and Equation \eqref{eq:motion-cartesian}, respectively. The two ion components, warm and hot, have densities $n_w = 0.93 n_0$ and $n_{h, \mathrm{eq}} = 0.07 n_0$, respectively, where $n_0$ is the reference plasma density. The initial warm ion distribution is an isotropic Maxwellian with a thermal velocity $v_{T, w} / v_{\mathrm{A}} = 0.156$, representing ions of ionospheric origin. The hot ions are initialized as a bi-Maxwellian, representing the injected ion population from the plasma sheet. They have perpendicular and parallel thermal velocities at the equator $v_{T\perp h, \mathrm{eq}} / v_{\mathrm{A}} = 1.73$ and $v_{T\parallel h, \mathrm{eq}} / v_{\mathrm{A}} = 0.707$, respectively, which gives an anisotropy $A = v_{T\perp h, \mathrm{eq}}^2 / v_{T\parallel h, \mathrm{eq}}^2 - 1 = 5$. This anisotropy is higher than the typically observed values \cite{yue2019relationship},\add{ leads to larger saturation amplitude, and may even alter the wave dispersion relation.} \change{and}{Nonetheless, such anisotropy} is chosen to reduce the saturation time, thereby lowering the computational cost. Because of this anisotropy, the thermal velocities and density of hot ions are adiabatically mapped to higher latitudes as\,
\begin{linenomath*}
\begin{equation}\label{eq:mapping-anisotropy-density}
    \begin{split}
    v_{T \parallel h} (z) &= v_{T \parallel h, \mathrm{eq}} ,\\
    v_{T \perp h} (z) &= v_{T \perp h, \mathrm{eq}} \left[1 + A \left(1 - \frac{1}{1 + \xi z^2}\right) \right]^{-\frac{1}{2}} ,\\
    n_h (z) &= n_{h, \mathrm{eq}} \left[1 + A \left(1 - \frac{1}{1 + \xi z^2}\right) \right]^{-1} .
    \end{split}
\end{equation}
\end{linenomath*}
The Maxwellian electrons act as a warm fluid for EMIC wave generation, and have an initial thermal velocity $v_{T, e} / v_{\mathrm{A}} = 4.69$. Particles are reflected back to the system when they strike the boundary. The simulation box is large enough so that the waves do not reach the boundary during the time span of the simulation.\add{ Considering the limitations of PIC simulations (e.g., small ion-to-electron mass ratio, high ion anisotropy), it is necessary to note that the simulated wave packets may not fully capture certain features of the observed wave packets.}

% why low noise delta-f PIC is needed
Since the nonresonant scattering rate decays exponentially away from resonance, a high signal-to-noise ratio is needed to distinguish the nonresonant scattering from the ``apparent'' spread of pitch angle caused by the particle noise. Otherwise, the spread of pitch angle away from resonance may be dominated by the particle noise. To this end, the low-noise $\delta f$ method is implemented in our simulations \cite<>[]{parker1993fully,denton1995deltaf,sydora2003low,tao2017investigations}. In the $\delta f$-PIC method, a weight $w = \delta f / f$ is assigned to each particle, where $f$ is the total distribution function of a species, and $\delta f = f - f_0$ is the difference between $f$ and the equilibrium distribution $f_0$. Besides the standard steps in a PIC loop, after every particle push, the weight is updated using
\begin{linenomath*}
\begin{equation}
    \frac{\mathrm{d} w}{\mathrm{d} t} = - (1 - w) \frac{1}{f_0} \left[\frac{q}{m}\left(\delta \mathbf{E} + \frac{1}{c} \mathbf{v} \times \delta \mathbf{B}\right) \cdot \frac{\partial f_0}{\partial \mathbf{p}}\right] .
\end{equation}
\end{linenomath*}
In the deposition of current density, the contribution from each particle is multiplied by its weight $w$. Because the same number of particles are used to sample the perturbed distribution $\delta f$ instead of the full distribution $f$, the perturbed distribution $\delta f$ is very well sampled, yielding a better statistical representation of the distribution function. The discrete particle noise, which scales as $\sqrt{\langle w^2 \rangle} / \sqrt{N}$ ($N$ beging the number of particles), is substantially reduced. The implementation details of the $\delta f$-PIC method are elaborated in \ref{append-deltaf}.

\subsection{Electron scattering by EMIC wave packets}
\begin{figure}[tphb]
    \centering
    \includegraphics[width=\textwidth]{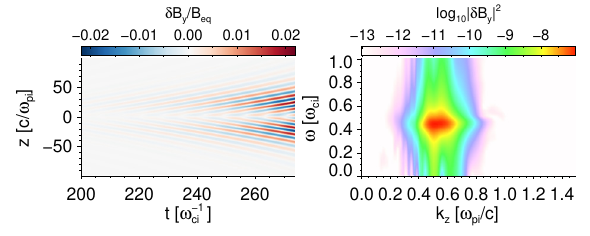}
    \caption{EMIC waves in the $\delta f$-PIC simulation. (a) Two EMIC wave packets generated at the equaotor propgagating to higher latitudes in the spatiotemporal domain. (b) EMIC wave power in the $\omega$-$k_z$ space.}
    \label{fig:propagation}
\end{figure}

\begin{figure}[tphb]
    \centering
    \includegraphics[width=\textwidth]{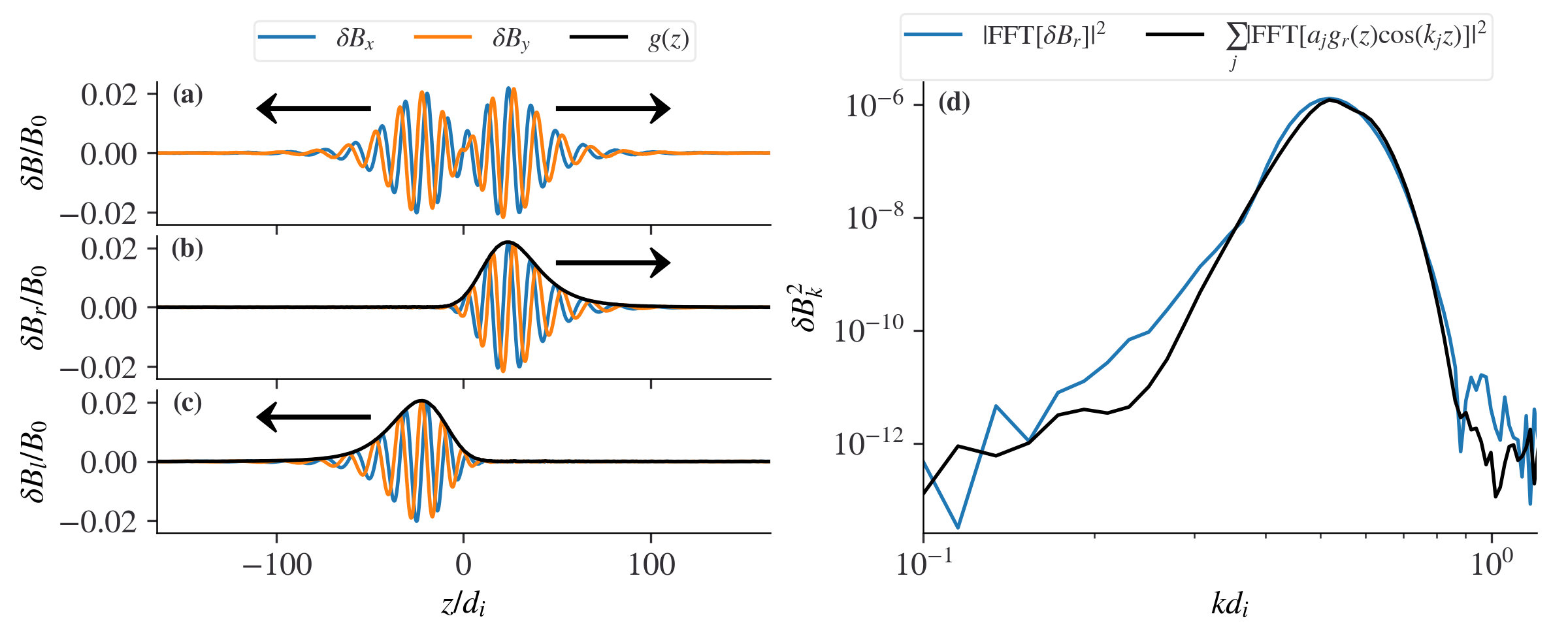}
    \caption{EMIC wave packets from the $\delta f$-PIC simulation. (a) Two EMIC wave packets propagating towards higher latitudes. (b) The northward propagating wave packet $\delta B_r$ has the right-handed magnetic helicity ($\delta B_x$ leads $\delta B_y$ by $90^\circ$ in space). (c) The southward propagating wave packet $\delta B_l$ has the left-handed magnetic helicity ($\delta B_y$ leads $\delta B_x$ by $90^\circ$ in space). The shape functions for $\delta B_r$ and $\delta B_l$, denoted as $g_r(z)$ and $g_l(z)$, are shown as the black curves in panels (b) and (c), respectively. (d) Magnetic power spectrum of $\delta B_r$ and the fitted spectrum as a function of wavenumber. The fitted spectrum is a Fourier transform of $\sum_{j=1}^{2} a_j g_r(z) \cos(k_j z)$ with $a_1 = 0.8$, $a_2 = 0.5$, $k_1 = 0.52 d_i^{-1}$, and $k_2 = 0.6 d_i^{-1}$. The magnetic power spectrum of $\delta B_l$ and its fitting are approximately the same as that of $\delta B_r$.}
    \label{fig:polarization}
\end{figure}

% short EMIC wave packets generated; separate left-handed and right-handed helicities; approximate power spectrum by (shape function * cosine)
The location for the fastest EMIC wave growth is the equatorial plane, where both the concentration and anisotropy of hot ions are maximized [see Equation \eqref{eq:mapping-anisotropy-density}]. Figure \ref{fig:propagation} shows two EMIC wave packets generated around the equator propagating towards higher latitudes and their power distribution in the $\omega$-$k_z$ space in the $\delta f$-PIC simulation. The propagation speed of these EMIC waves is around the Alfv\'en velocity $v_\mathrm{A} = c \omega_{ci} / \omega_{pi}$. The peak amplitude of EMIC waves is $\delta B_{\max} / B_{\mathrm{eq}} = 0.02$. The magnetic noise level in this case is $\delta B_{\mathrm{noise}} / B_{\mathrm{eq}} \approx 3 \times 10^{-8}$, which is small enough to not overshadow the nonresonant scattering. A snapshot of EMIC wave packets at $t = 274\, \omega_{ci}^{-1}$ is shown in Figure \ref{fig:polarization}(a). The dominant wavenumber of the EMIC wave packets is $0.52\, d_i^{-1}$ [Figure \ref{fig:polarization}(d)].

Because the two electron resonant interactions with the northward (the $+z$ direction) and southward (the $-z$ direction) propagating waves occur in two separate halves of one bounce period, the electron gyrophases during the two interactions are uncorrelated. In the evaluation of pitch angle diffusion rate, the northward and southward propagating waves can be treated as independent scatterers. For this reason, we separate the northward and southward propagating waves based on their right-handed ($\delta B_r$) and left-handed ($\delta B_l$) magnetic helicities, respectively \cite{terasawa1986decay} [Figures \ref{fig:polarization}(b) and \ref{fig:polarization}(c)]. The Fourier spectra of $\delta B_r$ and $\delta B_l$ are fitted by a superposition $\sum_j a_j g_{r, l}(z) \cos(k_j z)$ [Figure \ref{fig:polarization}(d)], where $k_j$ and $a_j$ are the wavenumber and fitting coefficient of the $j$th wave, respectively. This representation of the waveform is useful in the calculation of the scattering factor [Equation \eqref{eq:scattering-factor-approx}].

% setup of test particle simulations
Test particle simulations are performed using the EMIC wave packets from the $\delta f$-PIC simulation. The electron plasma frequency to electron gyrofrequency is $\omega_{pe}/\omega_{ce} = 18.2$, which corresponds to an electron minimum resonance energy $E_{\min} = 1.85$\,MeV. To evaluate the pitch angle diffusion rate at a given pitch angle and energy, an ensemble of $10^6$ electrons are initialized at $z/d_i = -150$ (well outside the EMIC wave packets) and move in the $+z$ direction. These electrons have the same initial equatorial pitch angle $\alpha_{\mathrm{eq}}$ and the same initial kinetic energy $E_k$ (scanned from $1$ to $1.85$\,MeV), and are uniformly distributed in gyrophase. Ensemble electron statistics are collected when electrons return to their initial position. Note that because of the different pitch angles resulted from scattering, the electrons may arrive at their initial position at slightly different times. The pitch angle diffusion rates are calculated for $18$ initial energies from $1$ to $1.85$\,MeV at the fixed initial pitch angle $10^\circ$.

\begin{figure}[tphb]
    \centering
    \includegraphics[width=0.6\textwidth]{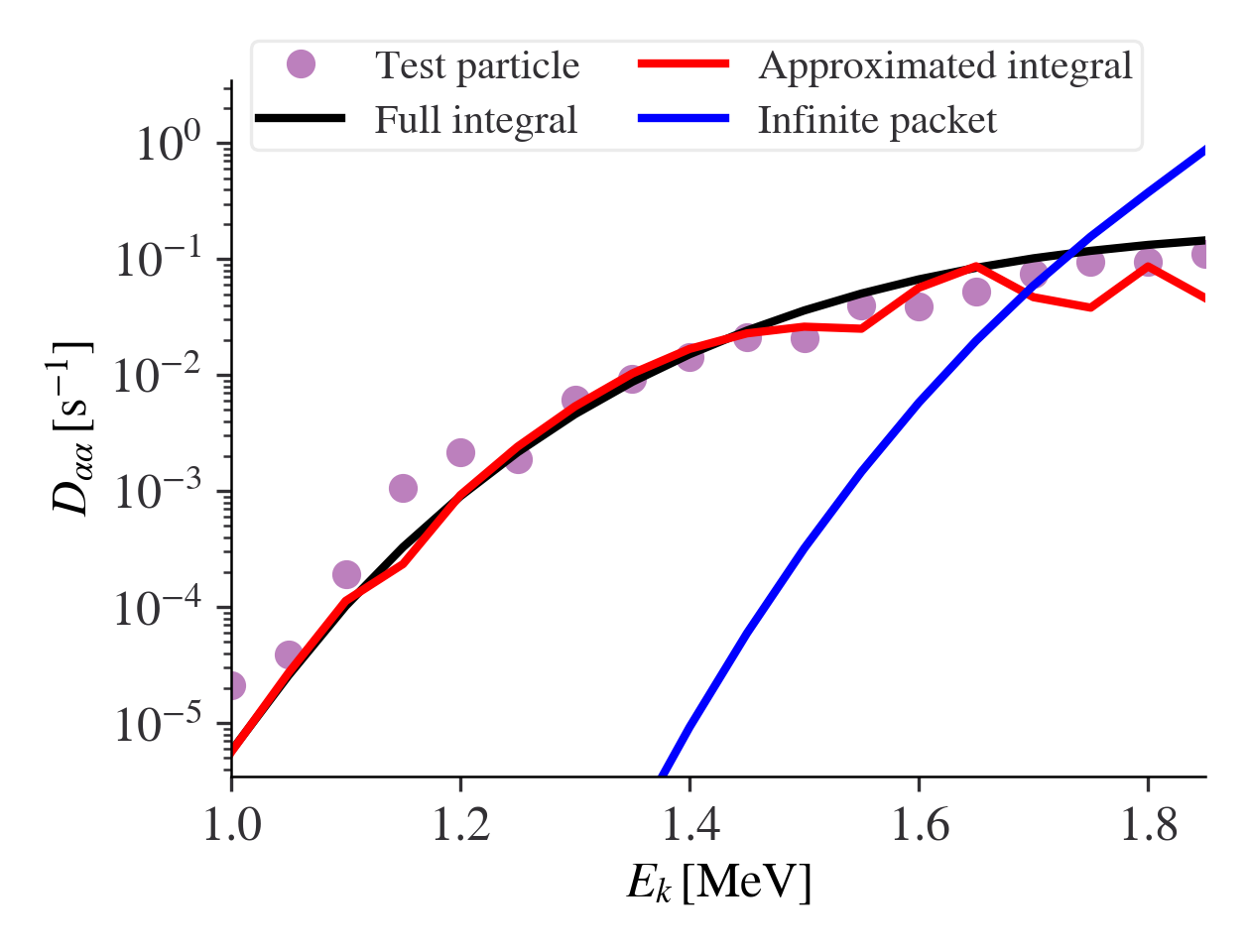}
    \caption{Comparison of pitch angle diffusion rate as a function of energy between theory and test particle simulations. Test particle results are shown in magenta dots. Theoretical predictions of $D_{\alpha \alpha}$ with the three versions of scattering factor from Equations \eqref{eq:G-exact}, \eqref{eq:scattering-factor-approx}, and \eqref{eq:G-long-limit} are shown in black, red, and blue curves, respectively. The nonsmooth behavior of the approximated integral (red) results from the nonsmooth Fourier spectrum $\hat{g}(\kappa)$ as seen in Figure \ref{fig:polarization}(d). The blue curve is only for the single frequency $\omega = 0.52\, \omega_{ci}$ of the carrier wave.}
    \label{fig:scattering-pic}
\end{figure}

% comparison of simulation results to theory
Figure \ref{fig:scattering-pic} shows a comparison of the pitch angle diffusion rate between test particle simulations and theoretical predictions. The predicted pitch angle diffusion rate from Equation \eqref{eq:Daa-theory} is calculated using three versions of the scattering factor $G$\change{:}{,} the full integral of Equation \eqref{eq:G-exact}, the approximate integral of Equation \eqref{eq:scattering-factor-approx}, and Equation \eqref{eq:G-long-limit} in the limit of an infinitely long wave packet. Both the full and approximate integrals capture the exponential decay of $D_{\alpha \alpha}$ below the minimum resonance energy. Evaluating $D_{\alpha \alpha}$ using the approximate integral is computationally more efficient than the full integral, because the phase integral is carried out analytically at the expense of sacrificing a small degree of accuracy due to the Taylor expansion. By comparing the test particle simulations with the electron scattering by an infinite wave, the realistic wave packets from $\delta f$-PIC simulations extend $D_{\alpha \alpha}$ from $\sim 10^{-1}\,\mathrm{s}^{-1}$ at $1.85$\,MeV to $\sim 10^{-3}\,\mathrm{s}^{-1}$ at $1.2$\,MeV, whereas the infinite wave packets make $D_{\alpha\alpha}$ exponentially decay to $\sim 10^{-3}\,\mathrm{s}^{-1}$ at $1.5$\,MeV. Because the fast variations of the wave shape $g(z)$ distribute the power around the wavenumber of the carrier wave [Figure \ref{fig:polarization}(d)], the power spread toward higher wavenumbers lowers the ``effective'' minimum resonance energy. It is worthy to note that the quasi-linear resonant diffusion has a hard cutoff of electron scattering at the minimum resonance energy $1.85$\,MeV, whereas the quasi-linear nonresonant diffusion extends the lower-bound energy of efficient scattering to $\sim 1.2$\,MeV (lowering the minimum resonance energy by $\sim 30\%$).

\section{Comparison with ELFIN observations}\label{sec:obserations}

%\toAnton{we may show several ELFIN events with EMIC-driven precipitations (no need to consider conjugated EMIC measurements) and estimate the energy range of precipitation outside the resonant precipitations that are characterized by precipitating/trapped flux ratio $\sim 1$. The we may show that nonresonant effects allow to cover this energy range..would be sufficient, i believe.}

To illustrate the effect on nonresonant electron scattering and verify the main conclusions of our analysis regarding this scattering process, we use the dataset from the ELFIN mission \cite{Angelopoulos20:elfin}. \add{Its two spinning CubeSats, A and B, move along low-altitude ($\sim 450$\,km) orbits and measure energy and pitch-angle distributions of energetic ($50-4000$\,keV) electrons with a time resolution of $1.5$\,seconds (half spin). For each half spin ELFIN covers the entire pitch-angle range: we evaluate a locally trapped flux by integrating within the pitch-angle range outside of the bounce loss cone, whereas a precipitating flux is obtained by integrating inside the bounce loss cone. Both data products, trapped and precipitating fluxes, are thus available with $1.5$\,second resolution.}\add{ The bounce loss cone is determined using the in-situ measured magnetic field and the magnetic field at $100$\,km altitude (at which electrons are considered to be precipitated in the upper atmosphere) from the International Geomagnetic Reference Field (IGRF) model }\cite{alken2021international}\add{.} Comparisons between locally trapped and precipitating electron fluxes, $j_{\perp}(E)$ and $j_{\parallel}(E)$, provide insight into different electron scattering mechanisms \cite<see discussion of different patterns of electron precipitation events in>[]{Mourenas21:jgr:ELFIN,angelopoulos2023energetic}. Here the focus is on those precipitation events with typical signatures of EMIC-driven electron precipitation. That is, $j_\parallel/j_\perp$ shows a peak at $>1$\,MeV, and $j_\parallel/j_\perp$ at $<500$\,keV is significantly smaller than $j_\parallel/j_\perp$ at $>1$\,MeV \cite<see the detailed analysis of such EMIC events in>[]{Grach22:elfin,angelopoulos2023energetic}. These conditions exclude from consideration all whistler-driven electron precipitation events, since those are characterized by decreasing $j_\parallel/j_\perp$ with increasing energy \cite<see examples in>[]{Tsai22,Chen22:microbursts}. We also exclude events having signatures of plasma sheet electron precipitation at the so-called isotropy boundary \cite<see examples in>[]{Wilkins23:arXiv,Artemyev22:jgr:ELFIN&THEMIS}, which lies at the interface between the plasma sheet and the outer radiation belt \cite{Sergeev83, Sergeev93:test}.

\begin{figure}[tphb]
\centering\includegraphics[width=\textwidth]{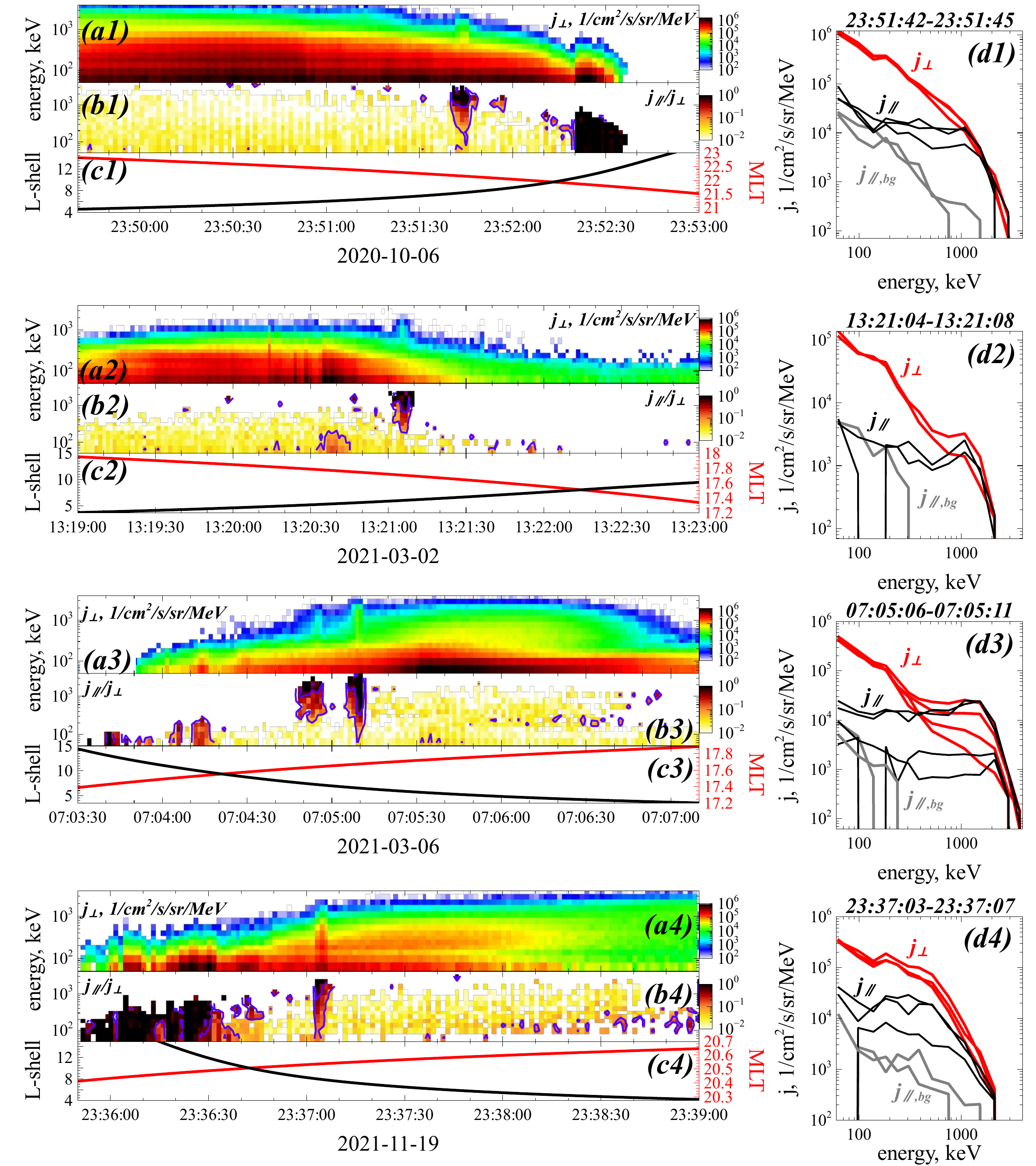}
\caption{Four examples of ELFIN observations of EMIC-driven electron precipitation events. Each panel is indexed by $(\alpha k)$, where $\alpha$ = $(a, b, c, d)$ indicates the panel number of a specific event, and $k$ is the event number. (a) Trapped fluxes. (b) The precipitating-to-trapped flux ratio $j_\parallel / j_\perp$. The blue lines show $0.05$ and $0.5$ contours of $j_\parallel / j_\perp$. (c) $L$-shell and MLT of ELFIN orbit. (d) The energy profiles of trapped, precipitating, and background fluxes. The intervals of EMIC-driven precipitation events are shown at the top of panels (d1)--(d4). Different lines with the same color indicates electron fluxes of different spins, which signify the electron flux variability from spin to spin.
\label{fig:elfin1}}
\end{figure}

Figure \ref{fig:elfin1} shows four examples of typical EMIC-driven precipitation events. All four events are located at the dusk flank, where one population of EMIC waves are usually detected [the other population being located in the dayside outer magnetosphere due to solar wind compression.] \cite<See>[]{Meredith14,Zhang16:grl,Jun19:emic,Keika13,Usanova12,ross2021variability}. % In this region, we do not expect a sharp flux gradient at the boundary of the plasma sheet and outer radiation belt, which is associated with the isotropy boundaries at the midnight \cite{Wilkins23:arXiv}. 

Panels (a1)--(a4) show the energy spectra of locally trapped electron population. The electron energies, as well as the electron flux magnitude, increase toward lower $L$-shells. At the times when the $\sim 300$\,keV electron flux becomes substantial $\sim 10^4$/cm$^2$\,/s/sr/MeV, ELFIN is considered crossing the plasma sheet-radiation belt boundary. 

Panels (b1)--(b4) show the precipitating-to-trapped flux ratio, $j_\parallel/j_\perp$. This ratio is generally small in the dusk flank [except for those electron precipitations from the plasma sheet, e.g., $>$23:51:20 in Panel (b1) and $<$23:36:30 in Panel (b4)], because of the absence of strong whistler-mode waves \cite{Agapitov13:jgr,Meredith12}, the main wave mode responsible for electron scattering into the loss-cone. However, all four events show time intervals of bursty precipitation with $j_\parallel/j_\perp \sim 1$ reaching the strong diffusion limit above $1$\,MeV \cite<e.g.,>[]{Kennel69}: 23:51:45 in Panel (b1), 13:21:05 in Panel (b2), 07:04:50 and 07:05:10 in Panel (b3), 23:37:05 in Panel (b4). These bursts of precipitation are most likely driven by EMIC waves, which scatter near-equatorial relativistic electrons with very high scattering rates \cite<e.g.,>[]{Summers&Thorne03, Albert03, Ni15}. An important property of these precipitation bursts is that $j_\parallel/j_\perp$ remains significantly larger than the background level [$(j_\parallel/j_\perp)_{\mathrm{bg}} \sim 0.1$] down to $\sim 100$--$300$\,keV. Such energies are generally well below the minimum resonance energy of EMIC waves \cite<see>[]{Kersten14,cao2017scattering,Chen19,Bashir22:grl}. Moreover, for $\sim 100-300$\,keV, the observed $j_\parallel/j_\perp$ is small (relative to $j_\parallel/j_\perp$ at $1$\,MeV) but still statistically significant, whereas the resonant scattering rates are expected to have a hard cutoff (i.e., drop to $0$) below a relatively high minimum resonance energy typically higher than $1$--$2$\,MeV \cite<e.g.,>[]{Summers&Thorne03,Albert03,Ni15}. Although the scattering of sub-relativistic ($<500$\,keV) electrons by EMIC waves may be explained by the presence of a population of small-amplitude EMIC waves with frequencies close to the proton cyclotron frequency \cite<e.g.,>[]{Zhang21,angelopoulos2023energetic}, ELFIN observations require that such a wave population must be present to explain the sub-relativistic precipitation for most events with simultaneous strong precipitation peaking above $1$\,MeV, which is not very probable. Thus, at least some portion of the sub-relativistic electron precipitation during EMIC-driven bursty precipitation events should be explained by nonresonant scattering. The ubiquity of amplitude modulations and short EMIC wave packets in observations \cite<e.g.,>[]{Usanova10,An22:prl} probably explains the presence of such higher wavenumber and higher frequency waves. The effects of such amplitude-modulated waves are directly taken into account in the present model of nonresonant scattering by wave packet edges.

Panels (d1)--(d4) zoom in on the precipitation bursts and show the energy spectra of trapped and precipitating fluxes during the bursts, as well as precipitating fluxes right before and after the bursts (i.e., the background level of precipitation). The energy range of the resonant interactions driving electron precipitation is characterized by $j_{\parallel} \approx j_{\perp}$. The spectra show such strong ($j_{\parallel} / j_{\perp} \approx 1$) precipitating fluxes in the range $>700$\,keV. Below this energy, the precipitating fluxes are still much higher than the background level of precipitation, but the scattering is not effective enough to provide $j_{\parallel} \approx j_{\perp}$. For the energy range $\in [100, 700]$\,keV, the precipitating fluxes depend only weakly on energy, whereas the trapped fluxes increase as energy decreases. This gives the overall trend of $j_{\parallel}/j_{\perp}$ going down with decreasing energy. 

%and the scattering rate $D_{\alpha\alpha} \propto (j_{\parallel}/j_{\perp})^2$ \cite{Kennel&Petschek66,Li13:POES} varies as $1/j^2_\perp \propto E^{2}$. \red{This scaling can be used to verify the proposed model of electron nonresonant scattering. [revise?]}

%To derive the $D_{\alpha\alpha}$ scaling with energy for the energy range below the minimum resonance energy, where $j_{\parallel}/j_{\perp}\approx 1$, we use the statistical dataset of ELFIN observations of EMIC-driven electron precipitation. This dataset includes one from \citeA{Angelopoulos22:arXiv}, but also adds \todo{???} events from 2022 ELFIN season. Figure \ref{fig:elfin2} shows averaged electron fluxes and flux ratio for this dataset. Electron losses \cite< precipitating fluxes minus backscattered fluxes, see details in>{Mourenas21:jgr:ELFIN} show almost the same spectrum as precipitating fluxes, so the latter one can be uses a good proxy of actual losses. The $j_{loss}/j_{\perp} \approx 1$ corresponding to the minimum resonance energy is around $1.5$MeV, the typical energy for electron scattering by EMIC waves \cite{Kersten14, Ni15}. Below this energy, $j_{loss}/j_{\perp}$ goes down as $j_{loss}/j_{\perp}\approx 0.065\times\gamma^2$ where $\gamma=E/m_ec^2+1$ \cite<this fitting coincides with one obtained for smaller ELFIN statistics, see>{Angelopoulos22:arXiv}. Therefore, we shall compare this scaling with theoretical results for $D_{\alpha\alpha}\propto (j_{\parallel}/j_{\perp})^2$.

\begin{figure}[tphb]
\centering\includegraphics[width=1\textwidth]{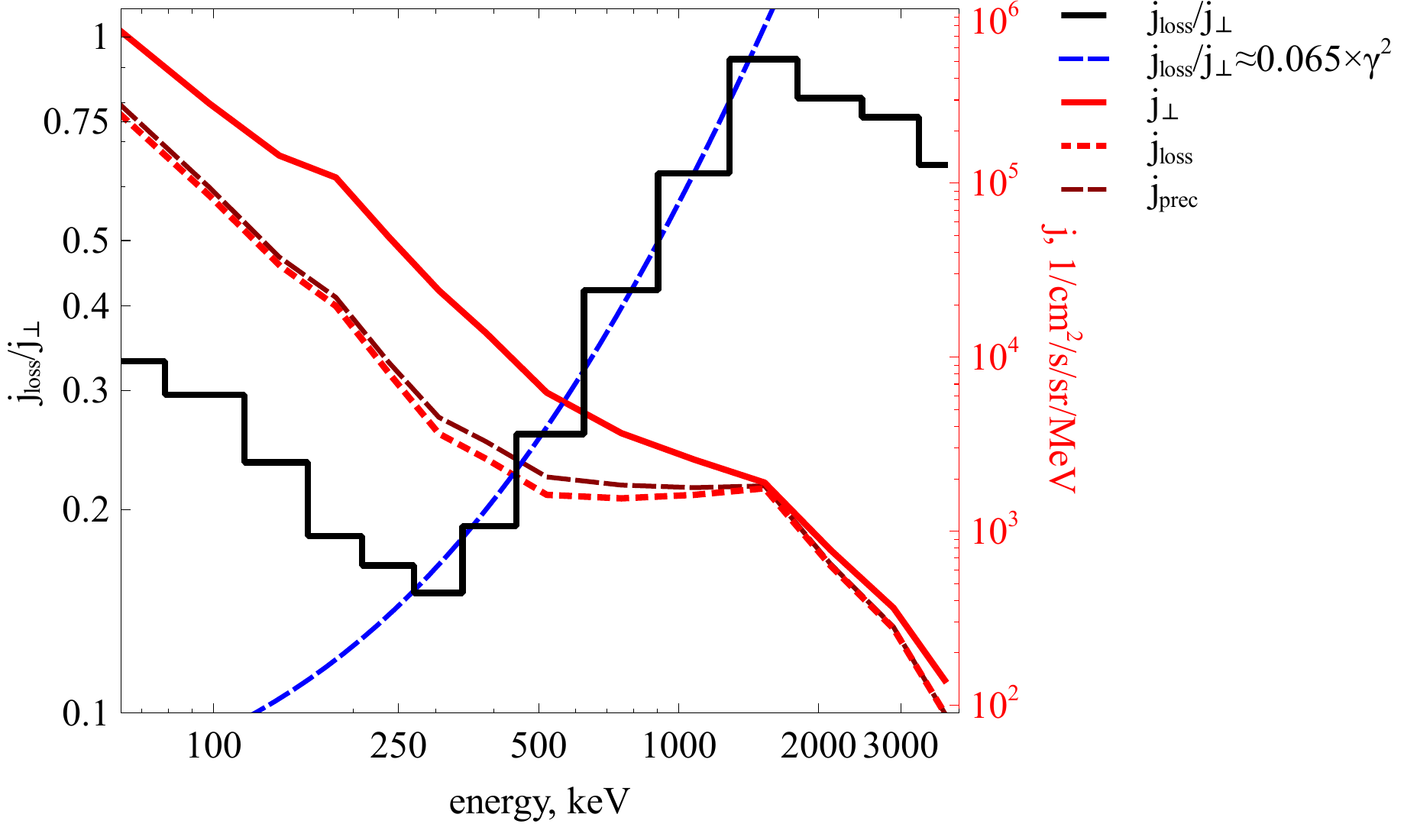}
\caption{Statistical observations of electron precipitation by ELFIN. The average fluxes for trapped, precipitating, and loss electrons are shown in solid red, dashed dark red, and dashed red, respectively. The electron loss flux is calculated by subtracting the back-scattered electron flux from the precipitating electron flux. The loss-to-trapped flux ratio and its fitting are shown in black and blue, respectively. This figure is replotted using the same dataset as analysed by \citeA{angelopoulos2023energetic}.
\label{fig:elfin2}}
\end{figure}

To gain further insight into the moderately efficient but statistically significant precipitation in the hundreds of keV range in the presence of strong precipitation at highly relativistic energies $>1$\,MeV, \add{the full statistical dataset ($\sim 180$ events with $\sim 500$ electron spectra)} of ELFIN observations of EMIC-driven electron precipitation\add{ is used} \cite<including the dataset in>[plus one more ELFIN season in year 2022]{angelopoulos2023energetic}. Figure \ref{fig:elfin2} shows the average (trapped, loss, and precipitating) electron fluxes, and the average precipitating-to-trapped flux ratio $\langle j_{\mathrm{prec}} / j_{\mathrm{trap}} \rangle$ of this dataset. The average electron loss flux \cite<the average precipitating flux minus the average back-scattered flux, see details in>[]{Mourenas21:jgr:ELFIN} shows almost the same spectrum as the average precipitating flux. So the latter can be used as a good proxy of actual losses. In the energy range $\in [0.3, 1.5]$\,MeV, the flux ratio rises from $\sim 0.15$ at $0.3$\,MeV to $\sim 1$ at $1.5$\,MeV, and can be approximately fitted as $\langle j_{\mathrm{prec}} / j_{\mathrm{trap}} \rangle \approx 0.065\times\gamma^2$. The flux ratio stays at a high value $\sim 0.75$ at energies $> 1.5$\,MeV. The flattening of $\langle j_{\mathrm{prec}} / j_{\mathrm{trap}} \rangle$ at energies above $1.5$\,MeV, as well as the falloff of $\langle j_{\mathrm{prec}} / j_{\mathrm{trap}} \rangle$ below $1.5$\,MeV, is consistent with the most common minimum cyclotron resonance energy being $>1$\,MeV \cite{Kersten14, Ni15}. The trend of the flux ratio at lower energies $<300$\,keV could be attributed to chorus-wave driven precipitation \cite<e.g.,>[]{Mourenas22:jgr:ELFIN}.

\begin{figure}[tphb]
    \centering
    \includegraphics[width=0.8\textwidth]{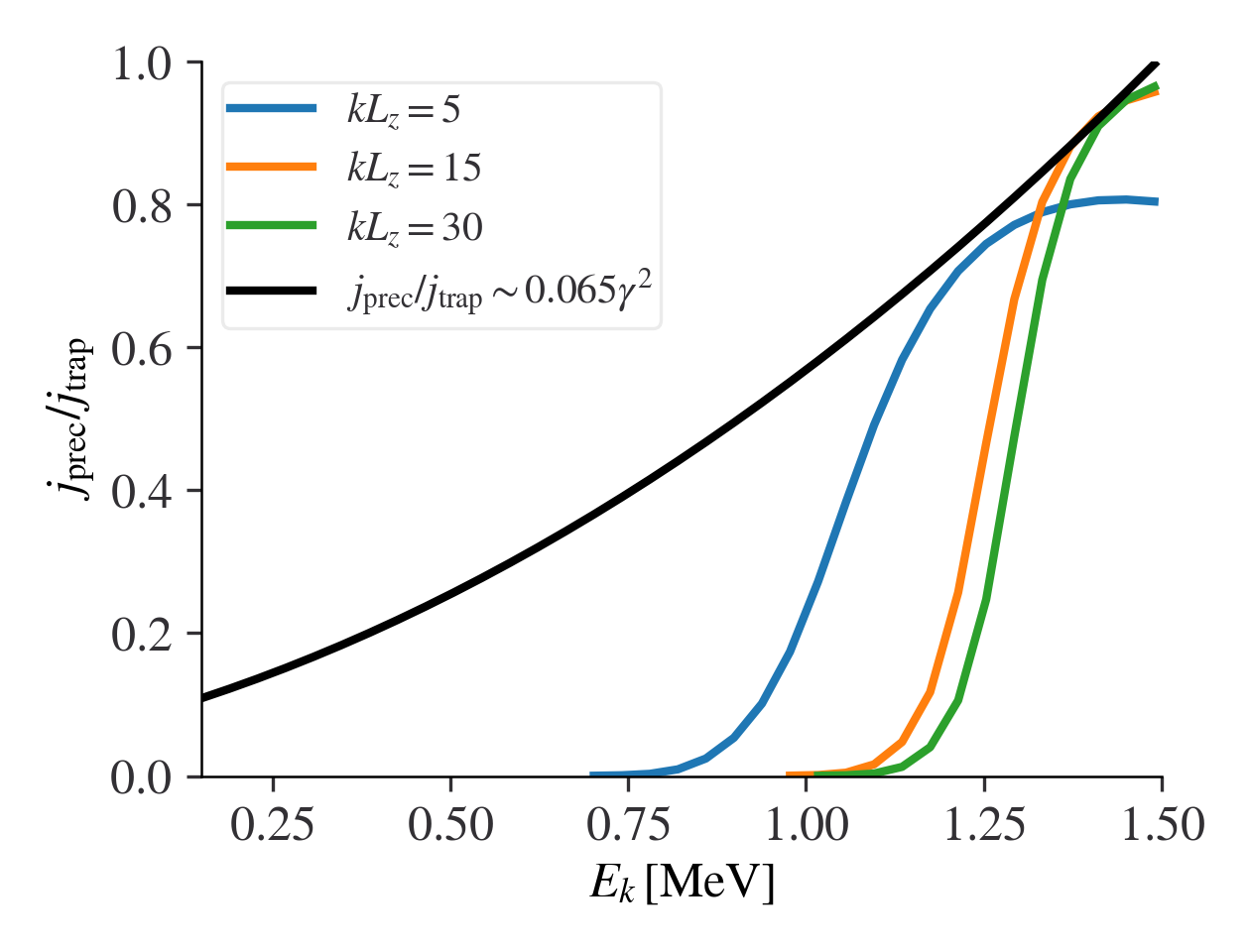}
    \caption{The precipitating-to-trapped flux ratio for a single frequency ($\omega / \omega_{ci}$ = $0.4$) wave packet with three different packet sizes $k L_z$ = $5, 15, 30$. The statistical measurements of electron precipitation from ELFIN, $j_{\mathrm{prec}} / j_{\mathrm{trap}} \approx 0.065 \gamma^2$, is plotted as a reference.}
    \label{fig:flux-ratio}
\end{figure}

To investigate how the proposed model of electron scattering by EMIC waves explains the precipitation spectrum, we first calculate the precipitating-to-trapped flux ratio for a given wave packet, taking nonresonant scattering into account.\add{ The relationship between the flux ratio at selected pitch-angles and the quasi-linear diffusion rate at the loss-cone angle is given by quasi-linear theory }\cite{Kennel&Petschek66,li2013constructing}. \add{Simply averaging the precipitating flux over pitch-angles yields }\cite{angelopoulos2023energetic}\add{:}\remove{ For $j_{prec}/j_{trap} \in [0.1,1]$, the flux ratio is related to the quasi-linear diffusion rate by}
\begin{linenomath*}
\begin{equation}\label{eq:flux-ratio}
    \frac{j_{\mathrm{prec}}}{j_{\mathrm{trap}}} \simeq \int_{0}^{1} \mathrm{d}x \frac{I_0(z_0 x) / I_0(z_0)}{1 + (z_0/20)I_1(z_0) / I_0(z_0)} ,
\end{equation}
\end{linenomath*}
where $I_0(\cdot)$ and $I_1(\cdot)$ are the modified Bessel functions of the first kind, $z_0 = 2 \alpha_{\mathrm{LC}} / \sqrt{D_{\alpha \alpha} \tau_b}$ measures the diffusion strength ($z_0 \ll 1$ and $z_0 \gg 1$ being the strong and weak diffusion, respectively), and $\alpha_{\mathrm{LC}}$ is the loss cone angle. Note that the precipitating flux $j_{\mathrm{prec}}$ has been averaged over the loss cone $\alpha < \alpha_{\mathrm{LC}}$. The trapped flux $j_{\mathrm{trap}}$ is measured at the equatorial pitch angle $\alpha_{\mathrm{trap}} = 1.05 \alpha_{\mathrm{LC}}$, providing the factor $\ln(\sin \alpha_{\mathrm{trap}} / \sin \alpha_{\mathrm{LC}}) \approx 1/20$ [see Equation (4.9) in \citeA{Kennel&Petschek66}]. This small factor implies that for a moderate diffusion rate $D_{\alpha \alpha} > (\alpha_{\mathrm{trap}} - \alpha_{\mathrm{LC}})^2 / (2 \tau_b) = \alpha_{\mathrm{LC}}^2 / (2\cdot 20^2 \tau_b)$, one has $z_0 = 2 \alpha_{\mathrm{LC}} / \sqrt{D_{\alpha \alpha} \tau_b} < 20\sqrt{2}$ and thus $j_{\mathrm{prec}} / j_{\mathrm{trap}} \sim 1/z_0 > 1/(20 \sqrt{2})$. The diffusion rate $D_{\alpha \alpha}$ is given by Equation \eqref{eq:Daa-theory}. We map between wave frequency and electron energy using the cyclotron resonance condition combined with the cold plasma dispersion relation. In this mapping, the energy of the maximum flux ratio ($\sim 1.5$\,MeV) from ELFIN statistical electron precipitation measurements corresponds to the frequency of the peak EMIC wave power ($\sim 0.4 \omega_{ci}$; $\omega_{ci}$ being the proton gyrofrequency) from Van Allen Probes statistical wave observations. It is first assumed below that the EMIC wave spectrum is a Dirac delta function with only one frequency $\omega = 0.4 \omega_{ci}$\add{ (note that frequency spectrum is used in this work for convenient comparisons with spacecraft observations, although wavenumber spectrum is the more directly relevant in the resonance condition)}. An ion composition of $>94\%$ protons is also assumed, as appropriate for when hydrogen band waves are present \cite{Kersten14,ross2022importance,angelopoulos2023energetic}. In this case, the mapping gives the wavenumber $k d_i = 0.52$ and the ratio of plasma frequency to electron gyrofrequency $\omega_{pe} / \omega_{ce} = 21.8$. For the purpose of demonstration, the wave amplitude $B_w / B_{\mathrm{eq}} = 0.005$ and the Gaussian wave shape $g(z) = e^{-z^2 / (2L_z^2)}$ are used. Figure \ref{fig:flux-ratio} shows the precipitating-to-trapped flux ratio for three packet sizes $k L_z = 5, 15, 30$. The value of $k L_z$ characterizes the number of wave periods in a wave packet. The falloff of the flux ratio below the minimum resonance energy is captured by considering nonresonant interactions. As the packet size decreases from $k L_z = 30$ to $k L_z = 5$, the lower-bound energy of significant precipitation extends from $\sim 1.25$\,MeV to $\sim 1$\,MeV. This figure shows how a single frequency (monochromatic) EMIC wave accounts for a narrow energy range of the precipitation burst, whereas the inclusion of nonresonant effects may extend the effective {\it energy width} of the precipitation burst.  

To account for the average $j_{\mathrm{prec}} / j_{\mathrm{trap}}$ seen in the ELFIN statistics, it is necessary to use an ensemble of wave packets. The probability distribution of wave packets, $P(l, B_w, \omega)$, is a function of packet size $l$, amplitude $B_w$, and frequency $\omega$ of peak wave power in the packet. Averaging the diffusion rate in Equation \eqref{eq:Daa-theory} over $(l, B_w, \omega)$, we obtain
\begin{linenomath*}
\begin{equation}\label{eq:Daa-theory-weighted}
    D_{\alpha\alpha} = \frac{e^2 \omega_{ce, \mathrm{eq}}}{4 (\gamma^2 - 1) m^2 c^4 \cos^2{\alpha_{\mathrm{eq}}} \omega_{ce}(z_c) \tau_b} \int \mathrm{d}\omega \int \mathrm{d}B_w \int \mathrm{d}l\, B_w^2 G(l, \omega) P(l, B_w, \omega) .
\end{equation}
\end{linenomath*}
On the one hand, one can construct $P(l, B_w, \omega)$ based on statistical wave measurements from equatorial spacecraft (such as Van Allen Probes), calculate $D_{\alpha \alpha}$ and further obtain $j_{\mathrm{prec}} / j_{\mathrm{trap}}$ to compare with statistical precipitation measurements from low-altitude spacecraft (such as ELFIN). This is a forward problem, requiring the empirical construction of $P(l, B_w, \omega)$. On the other hand, given the measured $j_{\mathrm{prec}} / j_{\mathrm{trap}}$, in principle, one can infer $P(l, B_w, \omega)$ by minimizing the cost function
\begin{linenomath*}
\begin{equation}\label{eq:cost-func}
	\mathcal{C}(l, B_w, \omega) = \sum_{i = 1}^{M} \left[ \left(\frac{j_{\mathrm{prec}}}{j_{\mathrm{trap}}}\right)_{\mathrm{theory}} (\gamma_i; l, B_w, \omega) - \left(\frac{j_{\mathrm{prec}}}{j_{\mathrm{trap}}}\right)_{\mathrm{measure}} (\gamma_i) \right]^2 = \sum_{i = 1}^{M} r_i^2 ,
\end{equation}
\end{linenomath*}
where $\gamma_i$ and $r_i$ are the Lorentz factor and the residual flux of the $i$th energy channel, respectively, and $M$ is the total number of energy channels. The theoretical $j_{\mathrm{prec}} / j_{\mathrm{trap}}$ is calculated by coupling Equations \eqref{eq:Daa-theory-weighted} to \eqref{eq:flux-ratio}. This is a nonlinear least square optimization problem, i.e., an inverse problem. In this study, we solve the inverse problem of inferring the statistical wave distribution based on the measured flux ratio. The forward problem will be treated in a separate study accompanied by a statistical analysis of spacecraft observations of EMIC waves providing $P(l, B_w, \omega)$. 

In the present inverse problem, it would still be difficult to find $P(l, B_w, \omega)$ without observational constraints from equatorial spacecraft, because of the large, three-dimensional parameter space $(l, B_w, \omega)$ and non-unique solution. For the purpose of demonstration, we assume $P(l, B_w, \omega) = \delta \left(l - L_z\right) \delta \left(B_w - B_w(\omega)\right)$, which simplifies the integral in Equation \eqref{eq:Daa-theory-weighted} as $\int \mathrm{d}\omega\, B_w^2(\omega) G(L_z, \omega)$. Thus, we need to find the magnetic power spectrum $B_w^2(\omega)$ for a given packet size $L_z$ so that the theoretical $j_{\mathrm{prec}} / j_{\mathrm{trap}}$ is as close to the measured $j_{\mathrm{prec}} / j_{\mathrm{trap}}$ as possible.

\begin{figure}[tphb]
    \centering
    \includegraphics[width=\textwidth]{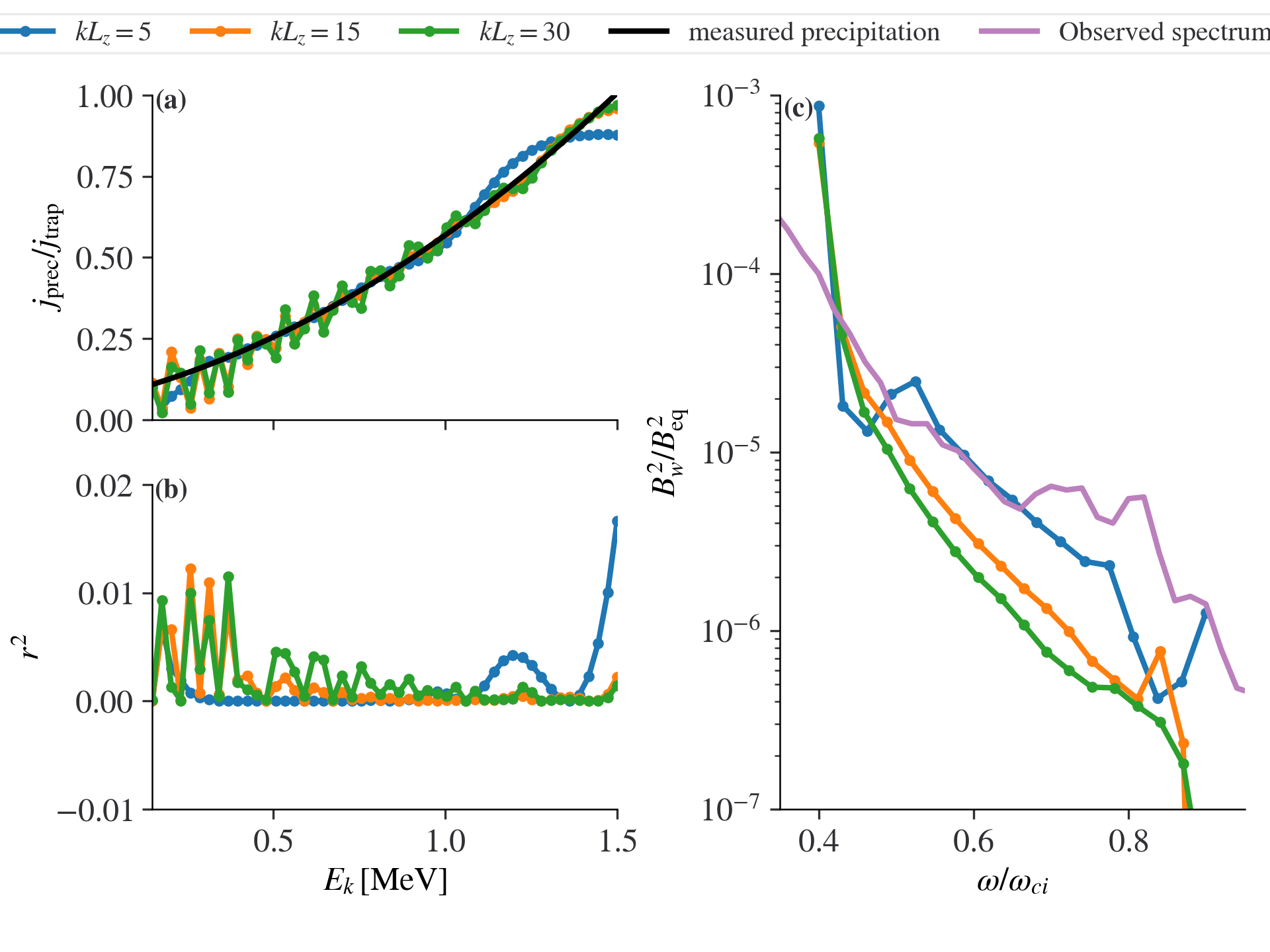}
    \caption{Inferred magnetic power spectra of EMIC waves and the resulting precipitating-to-trapped flux ratios. (a) The theoretical precipitating-to-trapped flux ratios for $k L_z = 5, 15, 30$. The theoretical ones are obtained by optimizing the wave power spectrum $B^2_w(\omega)$ so that the difference between the theoretical and measured precipitating-to-trapped flux ratios is minimized. The statistical measurements of electron precipitation from ELFIN, $j_{\mathrm{prec}} / j_{\mathrm{trap}} \approx 0.065 \gamma^2$, is plotted as a reference. (b) The residual between the theoretical and measured precipitating-to-trapped flux ratios. The residual is calculated using Equation \eqref{eq:cost-func}. (c) The optimized wave power spectra for $k L_z = 5, 15, 30$. The Van Allen Probes statistical observation of EMIC wave spectrum \cite<see Figure 19 from>[]{angelopoulos2023energetic} for MLT $12$--$16$ and the plasma-to-electron gyrofrequency ratio $\omega_{pe} / \omega_{ce} > 15$ is plotted as a reference. The observed wave power spectrum is normalized as $B_w^2 (\omega = 0.4 \omega_{ci}) / B_{\mathrm{eq}}^2 = 10^{-4}$. Note that the energy of maximum flux ratio $\sim 1.5$\,MeV from ELFIN statistics \cite{angelopoulos2023energetic} corresponds to the peak EMIC wave power at $\sim 0.4\, \omega_{ci}$ from Van Allen Probes statistics \cite{Zhang16:grl}. The EMIC frequencies that interact effectively with the energy range $<1.5$\,MeV are $> 0.4\, \omega_{ci}$.}
    \label{fig:optimization}
\end{figure}

In finding the optimized $B_w^2(\omega)$, unlike the one-to-one mapping between frequency and energy for resonant interactions, a single frequency is mapped to a range of energies for nonresonant interactions [Figure \ref{fig:flux-ratio}]. The Levenberg–Marquardt algorithm \cite{madsen2004methods,more1980user} is used to minimize the cost function [Equation \eqref{eq:cost-func}] and to find the optimized $B_w^2(\omega)$. The optimization details are given in \ref{append-optimize}. Figure \ref{fig:optimization} shows the inferred magnetic power spectra and the resulting precipitating-to-trapped flux ratio for three packet sizes $k L_z = 5, 15, 30$. The case of $k L_z = 30$ is an approximation for the limit of resonant interactions without strong wave modulations \cite<e.g.,>[]{An22:prl}. The strong precipitation near $1.5$\,MeV is caused by the dominant wave power at frequencies $\sim 0.4\,\omega_{ci}$, and the progressively weaker precipitation at lower energies is caused by higher frequencies with lower power. Comparing the results for different packet sizes, it is noted that, for shorter wave packets we may include a significant power for only the low-frequency part of the spectrum (almost monochromatic wave with $\sim 0.4\,\omega_{ci}$) and some small power (a factor of $<10^{-2}$) for the higher frequency part of the spectrum ($>0.5\,\omega_{ci}$), whereas for longer wave packets we should consider a wave spectrum smoothly decreasing away from the main frequency $\sim 0.4\,\omega_{ci}$ [which is consistent with Figure \ref{fig:flux-ratio}]. Based on the residual plot [Figure \ref{fig:optimization}(b)], shorter packets give a better agreement of the best fit with the observed $j_{\mathrm{prec}}/j_{\mathrm{trap}}$ at sub-MeV energies (due to the higher power present at higher frequency). Because the precipitation caused by nonresonant scattering for a lower frequency can overlap the precipitation caused by resonant scattering for a higher frequency, the effect of nonresonant scattering may be overshadowed by resonant scattering in this statistical picture. Nevertheless, some portion of the precipitation at sub-relativistic energies can be attributed to nonresonant scattering, which may be more significant in individual cases.

Comparing the observed wave spectrum for $\omega_{pe}/\omega_{ce} > 15$ with the inferred ones, the EMIC wave power needed to explain observations of electron precipitation at low energy via nonresonant scattering is lower than the average EMIC wave power observed in the 12--16 MLT range. This indicates that the spatially narrow wave packets (with significant wave power at high $k$ values) needed for nonresonant scattering of low-energy electrons are likely present only part of the time during each event.

\section{Conclusions}\label{sec:conclusions}
We study the nonresonant electron scattering by EMIC waves using a combination of theory, numerical simulations, and spacecraft observations. The main results are summarized as follows:
\begin{enumerate}
    \item The theoretical model of nonresonant scattering is verified for realistic wave-packets derived from self-consistent simulations.
    \item Using $\delta f$-PIC simulations, realistic EMIC wave packets are generated by hot, pitch-angle anisotropic ions through cyclotron resonant instability in a dipole field. By tracking test electrons moving through these wave packets, it is shown that significant nonresonant scattering occurs well below the resonance energies. The nonresonant scattering rate from test particle simulations agrees well with the theoretical model of nonoresonant scattering.
    \item The precipitating-to-trapped flux ratio is calculated including nonresonant scattering, and is compared to ELFIN observations. Shorter EMIC wave packets can provide a wider range of moderate precipitating fluxes below the resonance energies. Moreover, the wave power spectra parameterized by the packet size are inferred from the measured flux ratio by solving a nonlinear least square problem. With shorter wave packets, we can recover the precipitating fluxes below the resonance energies by using a significant power for the low-frequency part of the wave spectrum and a small but finite power (two orders of magnitude or more smaller than the peak wave power) for the high-frequency part of the wave spectrum.
\end{enumerate}
These results suggest that nonresonant scattering by strong, short EMIC wave packets can account for moderately intense precipitating fluxes below the minimum resonance energy. The formulation of nonresonant scattering can be used in radiation belt modeling to yield a more complete understanding of sub-relativistic electron precipitation driven by EMIC waves. Such modeling work should be preceded by a theoretical reconstruction of diffusion rates (at both resonance and nonresonance energies) from integrating a statistical distribution of wave packets \cite{shi2023properties} over packet sizes, amplitudes and frequencies, which is the next step for this line of work.

\appendix
\section{Calculation of the phase integral\label{append-calc-phase-integral}}
For resonant interactions, we calculate the phase integral along the contour in Figure \ref{fig:integral-contours}(a) as the following:
\begin{linenomath*}
\begin{equation}
	\int_{\Psi_l}^{\Psi_u} \frac{\mathrm{d}\Psi}{\dot{\Psi}(\tilde{z})} e^{i \Psi} = \frac{1}{(2 \ddot{\Psi}_0)^{\frac{1}{2}}} e^{i \Psi_0} \left(\int_{C_L} + \int_{C_0} + \int_{C_U}\right) \frac{\mathrm{d}\Psi}{(\Psi - \Psi_0)^{\frac{1}{2}}} e^{i (\Psi - \Psi_0)} .
\end{equation}
\end{linenomath*}
The integral around $\Psi_0$ (i.e., along the contour $C_0$) is $0$ as this contour becomes infinitesimally small. Transform to the new variable $\chi = \Psi - \Psi_0$ for the branch $C_L$, which gives $\chi = e^{-2\pi i} (\Psi - \Psi_0)$ for the branch $C_U$. Thus the integral is evaluated as
\begin{linenomath*}
\begin{equation}
	\begin{split}
		\int_{\Psi_l}^{\Psi_u} \frac{\mathrm{d}\Psi}{\dot{\Psi}(\tilde{z})} e^{i \Psi} &= \frac{1}{\left(2 \ddot{\Psi}_0\right)^{\frac{1}{2}}} e^{i \Psi_0} \left(\int_{\infty}^{0} \frac{\mathrm{d}\chi}{\chi^{\frac{1}{2}}} e^{i \chi} + \int_{0}^{\infty} \frac{\mathrm{d}\chi}{-\chi^{\frac{1}{2}}} e^{i \chi}\right) \\
		&= -\left(\frac{2}{\ddot{\Psi}_0}\right)^{\frac{1}{2}} e^{i \Psi_0} \int_{0}^{\infty} \frac{\mathrm{d}\chi}{\chi^{\frac{1}{2}}} e^{i \chi} \\
		&= -\left(\frac{8}{\ddot{\Psi}_0}\right)^{\frac{1}{2}} e^{i \Psi_0} \int_{0}^{\infty} \mathrm{d}y e^{iy^2} \\
		&= -\left(\frac{2\pi}{\ddot{\Psi}_0}\right)^{\frac{1}{2}} e^{i (\Psi_0+\frac{\pi}{4})} ,
	\end{split}
\end{equation}
\end{linenomath*}
where we have made the substitution $y = \sqrt{\chi}$ and used the Fresnel integral.

For nonresonant interactions, using the Cauchy integral theorem, we evaluate the phase integral along the path shown in Figure \ref{fig:integral-contours}(b):
\begin{linenomath*}
\begin{equation}
	\int_{\Psi_l}^{\Psi_u} \frac{\mathrm{d}\Psi}{\dot{\Psi}(\tilde{z})} e^{i \Psi} = -\frac{e^{i \Psi_0}}{(2 \ddot{\Psi}_0)^{\frac{1}{2}}} \left(\int_{C_1} + \int_{C_2} + \int_{C_U} + \int_{C_0} + \int_{C_L} + \int_{C_3} + \int_{C_4}\right) \frac{\mathrm{d}\Psi}{(\Psi - \Psi_0)^{\frac{1}{2}}} e^{i (\Psi - \Psi_0)} .
\end{equation}
\end{linenomath*}
The integral along $C_0$ is $0$ as the contour becomes infinitesimally small, and the integrals along $C_1$, $C_2$,  $C_3$, and $C_4$ are $0$ as the contours go to infinity. Transform to the new variable $i \chi = \Psi - \Psi_0$ for the branch $C_U$, which gives $i \chi = e^{2\pi i} (\Psi - \Psi_0)$ for the branch $C_L$. We have
\begin{linenomath*}
\begin{equation}
	\begin{split}
		\int_{\Psi_l}^{\Psi_u} \frac{\mathrm{d}\Psi}{\dot{\Psi}(\tilde{z})} e^{i \Psi} &= -\frac{1}{(2 \ddot{\Psi}_0)^{\frac{1}{2}}} e^{i \Psi_0} \left(e^{\frac{\pi i}{4}}\int_{\infty}^{0} \frac{ \mathrm{d}\chi}{\chi^{\frac{1}{2}}} e^{-\chi} +  e^{\frac{\pi i}{4}}\int_{0}^{\infty} \frac{ \mathrm{d}\chi}{-\chi^{\frac{1}{2}}} e^{-\chi}\right) \\
		&=  \left(\frac{2}{\ddot{\Psi}_0}\right)^{\frac{1}{2}} e^{i (\Psi_0 + \frac{\pi}{4})} \int_{0}^{\infty} \frac{ \mathrm{d}\chi}{\chi^{\frac{1}{2}}} e^{-\chi} \\
		&= \left(\frac{2 \pi}{\ddot{\Psi}_0}\right)^{\frac{1}{2}} e^{i (\Psi_0 + \frac{\pi}{4})} .
	\end{split}
\end{equation}
\end{linenomath*}
The result of this phase integral has the same form as that in the resonant case except for a phase factor $-1$.

\section{Implementation of the $\delta f$-PIC method \label{append-deltaf}}
The full Vlasov equation reads
\begin{linenomath*}
\begin{equation}
	\frac{\mathrm{d} f}{\mathrm{d} t} = \frac{\partial f}{\partial t} + \dot{\mathbf{x}} \cdot \frac{\partial f}{\partial \mathbf{x}} + \dot{\mathbf{p}} \cdot \frac{\partial f}{\partial \mathbf{p}} = 0 ,
\end{equation}
\end{linenomath*}
where $\mathbf{x}$ and $\mathbf{p}$ are the position and momentum coordinates, respectively, $\dot{\mathbf{x}} = \mathbf{v} = \mathbf{p} / (\gamma m)$ is the velocity, and $\dot{\mathbf{p}}$ denotes the force on the particles. In the $\delta f$ method, the total phase space density is separated as
\begin{linenomath*}
\begin{equation}
	f = f_0 + \delta f ,
\end{equation}
\end{linenomath*}
where $f_0$ and $\delta f$ are the equilibrium and perturbed distribution functions, respectively. Note that the perturbed part does not necessarily have to be small. The equilibrium part is consistent with the electromagnetic fields ($\mathbf{E}_0$, $\mathbf{B}_0$) and satisfies
\begin{linenomath*}
\begin{eqnarray}
	\partial f_0 / \partial t &=& 0, \\
	\mathbf{v} \cdot \frac{\partial f_0}{\partial \mathbf{x}} + \dot{\mathbf{p}}_0 \cdot \frac{\partial f_0}{\partial \mathbf{p}} &=& 0 ,\label{eq:f0-deriv}
\end{eqnarray}
\end{linenomath*}
where  $\dot{\mathbf{p}}_0$ is determined by the equilibrium fields
\begin{linenomath*}
\begin{equation}
	\dot{\mathbf{p}}_0 = \frac{q}{m} \left(\mathbf{E}_0 + \frac{1}{c} \mathbf{v} \times \mathbf{B}_0\right) .
\end{equation}
\end{linenomath*}
The evolution of $\delta f$ is
\begin{linenomath*}
\begin{equation}\label{eq:d-deltaf-dt-intermediate}
	\frac{\mathrm{d} \delta f}{\mathrm{d} t} = - \frac{\mathrm{d} f_0}{\mathrm{d} t} = - \left(\mathbf{v} \cdot \frac{\partial f_0}{\partial \mathbf{x}} + \dot{\mathbf{p}} \cdot \frac{\partial f_0}{\partial \mathbf{p}}\right) .
\end{equation}
\end{linenomath*}
Here $(\mathbf{v}, \dot{\mathbf{p}})$ is along the exact orbits, determined by both the equilibrium and perturbed electromagnetic fields. Noting that $\dot{\mathbf{p}} = \dot{\mathbf{p}}_0 + \delta \dot{\mathbf{p}}$, we subtract Equation \eqref{eq:f0-deriv} from \eqref{eq:d-deltaf-dt-intermediate} and obtain
\begin{linenomath*}
\begin{equation}
	\frac{\mathrm{d} \delta f}{\mathrm{d} t} = - \frac{\mathrm{d} f_0}{\mathrm{d} t} = - \delta \dot{\mathbf{p}} \cdot \frac{\partial f_0}{\partial \mathbf{p}} ,
\end{equation}
\end{linenomath*}
where
\begin{linenomath*}
\begin{equation}\label{eq:dot-dp}
	\delta \dot{\mathbf{p}} = \frac{q}{m} \left(\delta \mathbf{E} + \frac{1}{c} \mathbf{v} \times \delta \mathbf{B}\right) .
\end{equation}
\end{linenomath*}

In the initialization of $\delta f$ method, aside from sampling particles' $(\mathbf{x}_0, \mathbf{p}_0)$ from $f_0$ as in full-$f$ simulations, we assign a weight $w_i = \delta f / f$ to the $i$-th particle, uniformly distributed at $[-\varepsilon, \varepsilon]$, where $\varepsilon$ is a small number (e.g., $\varepsilon = 10^{-5}$ in our simulations). The weight is updated as
\begin{linenomath*}
\begin{equation}\label{eq:dwdt}
	\frac{\mathrm{d} w_i}{\mathrm{d} t} = - (1 - w_i) \left(\delta \dot{\mathbf{p}} \cdot \frac{\partial \ln f_0}{\partial \mathbf{p}}\right)_{\mathbf{x} = \mathbf{x}_i, \mathbf{p} = \mathbf{p}_i, t}  ,
\end{equation}
\end{linenomath*}
where the identity $1/f = (1 - w_j)/f_0$ is used in deriving this equation, and $\delta \dot{\mathbf{p}}$ and $\partial \ln f_0 / \partial \mathbf{p}$ are evaluated at the exact particle orbits. We notice that: (1) $\delta \dot{\mathbf{p}}$ is evaluated using Equation \eqref{eq:dot-dp}; (2) $\partial \ln f_0 / \partial \mathbf{p}$ only needs to be computed once on appropriate $(\mathbf{x}, \mathbf{p})$ grids at the beginning of simulation, and its values at the exact particle orbits are interpolated from the grids as simulation proceeds. In the deposition of charge and current densities, the perturbed distribution is represented by
\begin{linenomath*}
\begin{equation}
	\delta f(\mathbf{x}, \mathbf{p}, t) = \sum_{i} w_i S(\mathbf{x} - \mathbf{x}_i) \delta(\mathbf{p} - \mathbf{p}_i) ,
\end{equation}
\end{linenomath*}
where $S(\cdot)$ is the particle shape function. The charge and current densities are then computed using the perturbed distribution, based on which the perturbed fields $\delta \mathbf{E}$ and $\delta \mathbf{B}$ are calculated using Maxwell's equations.

\section{Optimization of the wave power spectrum \label{append-optimize}}
We discretize the wave power spectrum at $N$ frequencies ($\omega_1, \omega_2, \cdots, \omega_N$) as $B_j^2 = \int_{\omega_j - \frac{\delta \omega}{2}}^{\omega_j + \frac{\delta \omega}{2}} \mathrm{d}\omega B_w^2(\omega = \omega_j)$, where $j = 1, 2, \cdots, N$ indexes over frequencies, and $\delta \omega$ is the frequency spacing between two adjacent frequencies. Because $B_j^2$ is constrained as $0 < B_j^2 / B_{\mathrm{eq}}^2 < 1$, we transform it to
\begin{linenomath*}
\begin{equation}\label{eq:param-x}
    x_j = \ln \left(\frac{B_j^2 / B_{\mathrm{eq}}^2}{1 - B_j^2 / B_{\mathrm{eq}}^2}\right) ,
\end{equation}
\end{linenomath*}
which is used to replace $B_j^2$ as the independent variable since $x_j$ is unconstrained (i.e., $-\infty < x_j < +\infty$). Once $x_j$ is determined, $B_j^2$ can be obtained through the logistic function:
\begin{linenomath*}
\begin{equation}
    \frac{B_j^2}{B_{\mathrm{eq}^2}} = \frac{1}{1 + e^{-x_j}} .
\end{equation}
\end{linenomath*}

We aim to minimize the cost function
\begin{linenomath*}
\begin{equation}
    \mathcal{C}(\mathbf{x}) = \sum_{i = 1}^{M} \vert r_i(\mathbf{x}) \vert^2 ,
\end{equation}
\end{linenomath*}
where the residual $r_i$ is a nonlinear function of $\mathbf{x}$ as defined in Equation \eqref{eq:cost-func}, and each component of the parameter vector $\mathbf{x}$ is given in Equation \eqref{eq:param-x}. The number of parameters $N$ (i.e., the number of wave frequencies in this application) is required to be less than the number of measurements $M$ (i.e., the number of energy channels in this application). In general, nonlinear least square problems do not admit closed-form solutions and must be solved through iterative methods. In each iteration, the nonlinear problem is approximated by a linear problem locally (e.g., through the Taylor expansion), which admits a closed-form solution. This solution to the linear problem is then used to update the estimate for the nonlinear problem. After a certain number of iterations, the estimate may converge to a local minimum (depending on the initial guess) of the cost function. Different iterative methods (e.g., Newton, Gauss-Newton, Levenberg–Marquardt) mainly differ in what the step size and direction are chosen to update the estimate \cite{madsen2004methods}, but the basic idea is the same.

Our initial guess is motivated by the observed spectrum, which may be approximately modeled as \cite{Zhang16:grl} 
\begin{linenomath*}
\begin{equation}\label{eq:initial}
    B_j^2 \propto \omega_j^{-3} .
\end{equation}
\end{linenomath*}
In each iteration of the Gauss-Newton method, the residual is linearized around the current estimate $\Bar{\mathbf{x}}$:
\begin{linenomath*}
\begin{equation}
    r_i(\mathbf{x}) \approx r_i(\Bar{\mathbf{x}}) + \mathbf{J}_i \cdot \pmb{\delta} ,
\end{equation}
\end{linenomath*}
where $\pmb{\delta} = \mathbf{x} - \Bar{\mathbf{x}}$ and $\mathbf{J}_i = \partial r_i / \partial \mathbf{x}\vert_{\mathbf{x} = \Bar{\mathbf{x}}}$. The $j$th element of $\mathbf{J}_i$ is
\begin{linenomath*}
\begin{equation}\label{eq:jacobian}
	\begin{split}
		J_{ij} &= \frac{\partial r_i}{\partial x_j} = \frac{\partial r_i}{\partial z_0} \frac{\partial z_0}{\partial D_{\alpha\alpha}} \frac{\partial D_{\alpha\alpha}}{\partial B_j} \frac{\partial B_j}{\partial x_j} \\
		&= -\frac{z_0}{2} \frac{D_{\alpha\alpha, j}}{D_{\alpha\alpha}} \left(1 - \frac{B_j^2}{B_{\mathrm{eq}^2}}\right) \\
  & \times \int_{0}^1 \mathrm{d}x \frac{x I_1(x z_0) \left[I_0(z_0) + \frac{z_0}{20}I_1(z_0)\right] - I_0(x z_0) \left[(\frac{21}{20}-\frac{z_0}{40})I_1(z_0) + \frac{z_0}{20}I_0(z_0)\right]}{\left[I_0(z_0) + \frac{z_0}{20} I_1(z_0)\right]^2} ,
	\end{split}
\end{equation}
\end{linenomath*}
where $D_{\alpha\alpha, j}$ is the diffusion rate contributed by the waves at frequency $\omega_j$, and $D_{\alpha \alpha}$ is the total diffusion rate. The minimization of the linearized residual $\sum_{i=1}^{M} \vert r_i(\Bar{\mathbf{x}}) + \mathbf{J}_i \cdot \pmb{\delta} \vert^2$ admits a closed-form solution
\begin{linenomath*}
\begin{equation}
    \pmb{\delta} = -(\mathbf{J}^T(\Bar{\mathbf{x}}) \mathbf{J}(\Bar{\mathbf{x}}))^{-1} \mathbf{J}^T(\Bar{\mathbf{x}}) r(\Bar{\mathbf{x}}) ,
\end{equation}
\end{linenomath*}
where $\mathbf{J} (\Bar{\mathbf{x}})$ and $r (\Bar{\mathbf{x}})$ are formed by stacking $\mathbf{J}_i (\Bar{\mathbf{x}})$ and $r_i (\Bar{\mathbf{x}})$, respectively. The current estimate $\Bar{\mathbf{x}}$ is then replaced by $\Bar{\mathbf{x}} + \pmb{\delta}$.

The Levenberg–Marquardt method is a modification of the Gauss-Newton method by adding a weighted diagonal term when solving the linear system $( \mathbf{J}^T(\Bar{\mathbf{x}}) \mathbf{J}(\Bar{\mathbf{x}}) + \lambda \mathbf{I} ) \pmb{\delta} = -\mathbf{J}^T(\Bar{\mathbf{x}}) r (\Bar{\mathbf{x}})$, where $\lambda$ is a weight to be adjusted according to a set of rules \cite<see details in >[]{more1980user}. This modification can provide robustness to the solution of nonlinear least square problems. To this end, we optimize the wave power spectrum using the Levenberg–Marquardt method from MINPACK (\url{https://netlib.org/minpack/}). The inputs to the program are the initial guess specified in Equation \eqref{eq:initial}, and the Jacobian matrix specified in Equation \eqref{eq:jacobian}. The program is stopped when the estimate $\Bar{\mathbf{x}}$ converges to a local minimum of the cost function.

\acknowledgments
This work was supported by NSF awards 2019914, 2108582, 2021749, and NASA awards 80NSSC20K0917, 80NSSC22K1634, 80NSSC20K1270, and 80NSSC23K0403. We are grateful to NASA's CubeSat Launch Initiative for ELFIN's successful launch in the desired orbits. We acknowledge early support of ELFIN project by the AFOSR, under its University Nanosat Program, UNP-8 project, contract FA9453-12-D-0285, and by the California Space Grant program. We acknowledge critical contributions of numerous volunteer ELFIN team student members. We would like to acknowledge high-performance computing support from Cheyenne (\url{doi:10.5065/D6RX99HX}) provided by NCAR's \citeA{cheyenne}, sponsored by the National Science Foundation. We would also like to acknowledge the OSIRIS Consortium, consisting of UCLA and IST (Lisbon, Portugal) for the use of \texttt{OSIRIS} and for providing access to the \texttt{OSIRIS} 4.0 framework.

\section*{Open Research} \noindent ELFIN data is available at \url{https://data.elfin.ucla.edu/}. Data analysis was done using SPEDAS V4.1 \cite{Angelopoulos19}, available at https://spedas.org/. The computation software and data are available at \url{https://doi.org/10.5281/zenodo.8122490} \cite{an_2023_8122490}.

%\bibliography{full,addon}

\end{document}